\documentclass[acmsmall,nonacm]{acmart}

\usepackage{tcolorbox}
\usepackage[utf8]{inputenc} 
\usepackage[T1]{fontenc}    
\usepackage{amsfonts}       
\usepackage{nicefrac}       
\usepackage{tabularx}
\usepackage{multirow}
\usepackage{xspace}
\usepackage{adjustbox}
\usepackage{tikz}
\usepackage{makecell}
\usepackage{wasysym}
\usepackage{enumitem}
\usepackage[normalem]{ulem}

\def\Snospace~{\S{}}

\usepackage{xstring}
\newcommand{\PP}[1]{
\vspace{2px}
\noindent{\bf \IfEndWith{#1}{.}{#1}{#1.}}
}
\newcommand{\squishlist}{
\begin{itemize}[noitemsep,nolistsep]
  \setlength{\itemsep}{-0pt}
}
\newcommand{\squishend}{
  \end{itemize}
}

\newcommand{\sys}[1]{AgentVault\xspace}

\newcommand{\dimtrust}{Input Trust\xspace}

\newcommand{\dimsensitivity}{Access Sensitivity\xspace}

\newcommand{\dimworkflow}{Workflow\xspace}
\newcommand{\dimaction}{Action\xspace}
\newcommand{\dimmemory}{Memory\xspace}
\newcommand{\dimtool}{Tool\xspace}
\newcommand{\diminterface}{User Interface\xspace}
\newcommand{\riskattack}{\hyperref[risk:attack-surface]{R1}}
\newcommand{\riskwronginst}{\hyperref[risk:wrong-instruction]{R2}}
\newcommand{\riskunsafedataflow}{\hyperref[risk:data-flow]{R3}}
\newcommand{\riskhallucination}{\hyperref[risk:hallucination]{R4}}
\newcommand{\riskconfidentiality}{\hyperref[risk:confidentiality]{R5}}
\newcommand{\riskintegrity}{\hyperref[risk:integrity]{R6}}
\newcommand{\riskavailability}{\hyperref[risk:availability]{R7}}

\newcommand{\vectorindirect}{\hyperref[vector:indirectinjection]{V1}}
\newcommand{\vectordata}{\hyperref[vector:datainjection]{V2}}
\newcommand{\vectortool}{\hyperref[vector:tool]{V3}}
\newcommand{\vectordirect}{\hyperref[vector:directinjection]{V4}}
\newcommand{\vectormodel}{\hyperref[vector:model]{V5}}
\newcommand{\vectormemory}{\hyperref[vector:memory]{V6}}

\acmJournal{CSUR}
\acmVolume{0}
\acmNumber{0}
\acmArticle{0}
\acmYear{2026}
\acmMonth{2}
\acmDOI{10.1145/0000000.0000000}

\copyrightyear{2026}
\setcopyright{none}

\title{The Attack and Defense Landscape of Agentic AI: A Comprehensive Survey}

\author{Juhee Kim}
\authornote{Work done during the author's visit to UC Berkeley.}
\affiliation{%
  \institution{UC Berkeley / Seoul National University}
  \city{Berkeley}
  \state{CA}
  \country{USA}
}
\email{kimjuhi96@snu.ac.kr}

\author{Xiaoyuan Liu}
\affiliation{%
  \institution{UC Berkeley}
  \city{Berkeley}
  \state{CA}
  \country{USA}
}
\email{xiaoyuanliu@berkeley.edu}

\author{Zhun Wang}
\affiliation{%
  \institution{UC Berkeley}
  \city{Berkeley}
  \state{CA}
  \country{USA}
}
\email{zhun.wang@berkeley.edu}

\author{Shi Qiu}
\authornotemark[1]
\affiliation{%
  \institution{UC Berkeley}
  \city{Berkeley}
  \state{CA}
  \country{USA}
}
\email{stephenqius@gmail.com}

\author{Bo Li}
\affiliation{%
  \institution{University of Illinois Urbana-Champaign}
  \city{Champaign}
  \state{IL}
  \country{USA}
}
\email{lbo@illinois.edu}

\author{Wenbo Guo}
\affiliation{%
  \institution{UC Santa Barbara}
  \city{Santa Barbara}
  \state{CA}
  \country{USA}
}
\email{henrygwb@ucsb.edu}

\author{Dawn Song}
\affiliation{%
  \institution{UC Berkeley}
  \city{Berkeley}
  \state{CA}
  \country{USA}
}
\email{dawnsong@cs.berkeley.edu}

\begin{document}

\begin{abstract}
AI agents that combine large language models with non-AI system components are rapidly emerging in real-world applications, offering unprecedented automation and flexibility. 
However, this unprecedented flexibility introduces complex security challenges fundamentally different from those in traditional software systems. 
This paper presents the first systematic and comprehensive survey of AI agent security, including an analysis of the design space, attack landscape, and defense mechanisms for secure AI agent systems. 
We further conduct case studies to point out existing gaps in securing agentic AI systems and identify open challenges in this emerging domain. 
Our work also introduces the first systematic framework for understanding the security risks and defense strategies of AI agents, serving as a foundation for building both secure agentic systems and advancing research in this critical area.
\end{abstract}



\received{March 2026}

\maketitle
\section{Introduction}
\label{sec:intro}

The rapid advancement of agentic AI systems has fundamentally transformed the AI landscape, marking a paradigm shift from isolated language models to integrated hybrid systems that combine Large Language Models (LLMs) with diverse software components.
These hybrid systems demonstrate unprecedented capabilities by seamlessly integrating AI reasoning and traditional software, enabling dynamic tool usage and autonomous task execution.
Agentic AI systems now power a wide range of applications, from a simple chatbot~\cite{openai-chatgpt,google-gemini} to software development~\cite{github-copilot,cursor,gemini-code-assist} and web browsing automation~\cite{comet-browser}.

The security implications of agentic AI systems are becoming increasingly critical, as recent incidents demonstrate the severity of their vulnerabilities.
Prompt injection attacks have been exploited to access private GitHub repositories~\cite{github-mcp-exploited}, while remote code execution vulnerabilities~\cite{cve-2024-5565,CVE-2025-54795} have enabled attackers to gain unauthorized system access.
Data exfiltration attacks~\cite{aim-echoleak,github-data-exfiltration} have compromised sensitive information through malicious document attachments and email forwarding, while servers have exposed user chat and credential data~\cite{march-20-chatgpt,langsmith-leak}. 
Attackers are also exploiting web agents to gain access to users' personal banking accounts~\cite{comet-prompt-injection}.
These incidents underscore that while the flexibility and automation capabilities make agentic systems powerful, they also create complex security challenges, which differ fundamentally from those associated with traditional software systems or standalone AI models.

Although existing research has made important contributions to understanding AI agent security, most efforts have focused on specific attack vectors or individual system components, mainly prompt injection attacks~\cite{liu2024formalizing,zhan24injecagent,yi2025benchmarking,zhang2025asb,beurer2025design}.
However, these studies lack a comprehensive perspective that considers the defense of agentic systems as a whole.
The community needs a systematic framework for understanding how the integration of multiple components introduces novel attack surfaces and demands fundamentally different security approaches.

This paper addresses this gap by providing the first comprehensive systematization of knowledge on the security landscape of agentic AI systems.
We approach agent security from a holistic systems perspective, examining how the combination of LLMs and traditional software creates unique security challenges that cannot be mitigated through component-level defenses alone.
Our analysis begins with a systematic characterization of agent design dimensions that influence security properties, followed by a comprehensive taxonomy of attack vectors and security risks across the entire agent ecosystem.
We then conduct a systematic survey of existing defense mechanisms, categorizing them based on their protection approaches and identifying critical gaps in current security strategies.
Finally, we conduct case studies on various real-world agents, including a concrete case of AutoGPT, to further highlight gaps in existing defenses for agentic systems.

Our contributions are threefold:
\begin{itemize}
\item \textbf{Agent Design Dimensions:} We present a systematic framework that characterizes agentic AI systems through seven key design dimensions: input trust, access sensitivity, workflow, action, memory, tool, and user interface. We then analyze how flexibility along each dimension impacts security risks and map these dimensions to established frameworks, including MITRE ATLAS and OWASP Top 10 for LLM.

\item \textbf{Comprehensive Attack Landscape and Taxonomy:} We develop a systematic taxonomy of attack vectors organized by threat models (i.e., external, user-level, and internal adversaries) and provide a comprehensive classification of seven security risk categories spanning the CIA triad, along with a system-level analysis of risk interactions and amplification patterns.

\item \textbf{Defense Landscape Systematization:} We systematically survey existing defense mechanisms and conduct various case studies, identifying specific design dimensions for defense and open challenges.

\end{itemize}

To the best of our knowledge, this is the first work to systematically analyze the security landscape of agentic AI systems from a comprehensive system perspective.
Our systematization provides a foundational framework for understanding security risks and defense strategies in agentic AI systems, guiding future research toward building secure agentic systems.
This work serves as a handbook for researchers and developers working with agentic AI systems.
\section{Overview}
\label{s:overview}

In this section, we define the scope of our work, introduce our methodology, and discuss the key differences from existing agent security surveys.

\PP{Scope}
We focus on security risks and defenses that are unique to, or significantly amplified in, agentic systems compared to traditional software and standalone LLMs.
We first characterize how agents differ from standalone models across seven design dimensions (\autoref{sec:agent}).
We then analyze agent-specific security risks both at the component and system levels (\autoref{sec:attack}).
For risks that exist in non-agentic LLMs~(e.g., jailbreak, hallucination), we emphasize how agent autonomy and environment access magnify their impact~(e.g., data exfiltration, unintended system manipulation).
We exclude attacks targeting model internals, such as model inversion~\cite{fredrikson2015model}, as these do not fundamentally worsen in agentic contexts where agents operate purely through inference-time input-output interfaces.
Finally, \autoref{sec:defense} systematizes the agent defense landscape, describing existing approaches and open challenges.

\PP{Methodology}
First, we define the agent structure and design dimensions based on existing definitions and real-world agent implementations. This forms the foundation of our agent risk taxonomy, which analyzes risks at both the component and system levels.
We also refer to the taxonomy of the OWASP Top 10 for LLM Applications~\cite{owasp-top-10-llm} and MITRE ATLAS Matrix~\cite{mitre-atlas}.
We then propose the defense landscape for agentic systems by applying the defense in-depth principle, informed by traditional system security.
Under this framework, we conduct a systematic review of academic literature and web documents on agentic AI security from 2023 to October 2025, corresponding to the proliferation of agent systems~\cite{yao2023react}. 
We search with keywords across four dimensions: i) general terminology (e.g., \emph{agent security}), ii) OWASP-defined risks (e.g., \emph{prompt injection}, \emph{memory poisoning}), iii) component-centric terms (e.g., \emph{RAG security}, \emph{tool security}), and iv) traditional security adaptations (e.g., \emph{isolation}, \emph{access control}).
We prioritize top-tier security venues (e.g., USENIX Security, IEEE S\&P, CCS, and NDSS) and ML conferences (e.g., NeurIPS, ICLR, and ICML), alongside high-impact preprints, industry whitepapers, and CVEs.
We manually exclude work on standalone models to focus explicitly on agents.
The review yields 128 papers, including 51 attack methods and 60 defense methods; the remaining works address both attacks and defenses or focus on case studies without introducing new methods.

\PP{Differences From Existing Surveys}
Existing surveys on AI and LLM security primarily emphasize model-level threats and overlook the expanded attack surface and downstream consequences introduced by agentic systems~\cite{grosse2024towards,liu2024formalizing,jia2025critical,wang2025sok}.
Prior agent-focused studies typically narrow their scope to specific attack vectors~\cite{zhang2025asb}, agent types~\cite{li2024personal,lee2025takedown}, or individual components~\cite{hou2025model}, while design-oriented work proposes high-level principles without systematizing attacks or defenses~\cite{zhang2025llm,beurer2025design}.
The closest surveys~\cite{deng2025ai,yu2025survey} consider cross-component interactions but largely limit their defense taxonomies to model-based techniques.
In contrast, our work provides a unified risk taxonomy covering all agent components and their interactions, together with a defense-in-depth framework that integrates both model-based and system-level defenses, offering actionable guidance for securing agentic systems.

\section{Design Landscape of Agentic AI Systems}
\label{sec:agent}

\subsection{Design Components}
\label{subsec:agent_define}

In this paper, we define AI agents as~\emph{hybrid software systems} that combine traditional software components with AI models.
An AI agent typically consists of the following components.

\PP{LLMs} As the \emph{brain} of the agent, the LLM(s) receive and analyze user tasks, create a step-by-step plan, and take actions following the plan.
These processes can be performed by a single or multiple LLMs, each tailored to a specific role.

\PP{Memory} Memory stores the agent’s internal knowledge base and historical action trajectories.
This information is often vectorized for efficient retrieval and can facilitate the agent to make future decisions based on knowledge and prior experience.

\begin{figure}[t]
    \centering
    \includegraphics[width=0.9\linewidth]{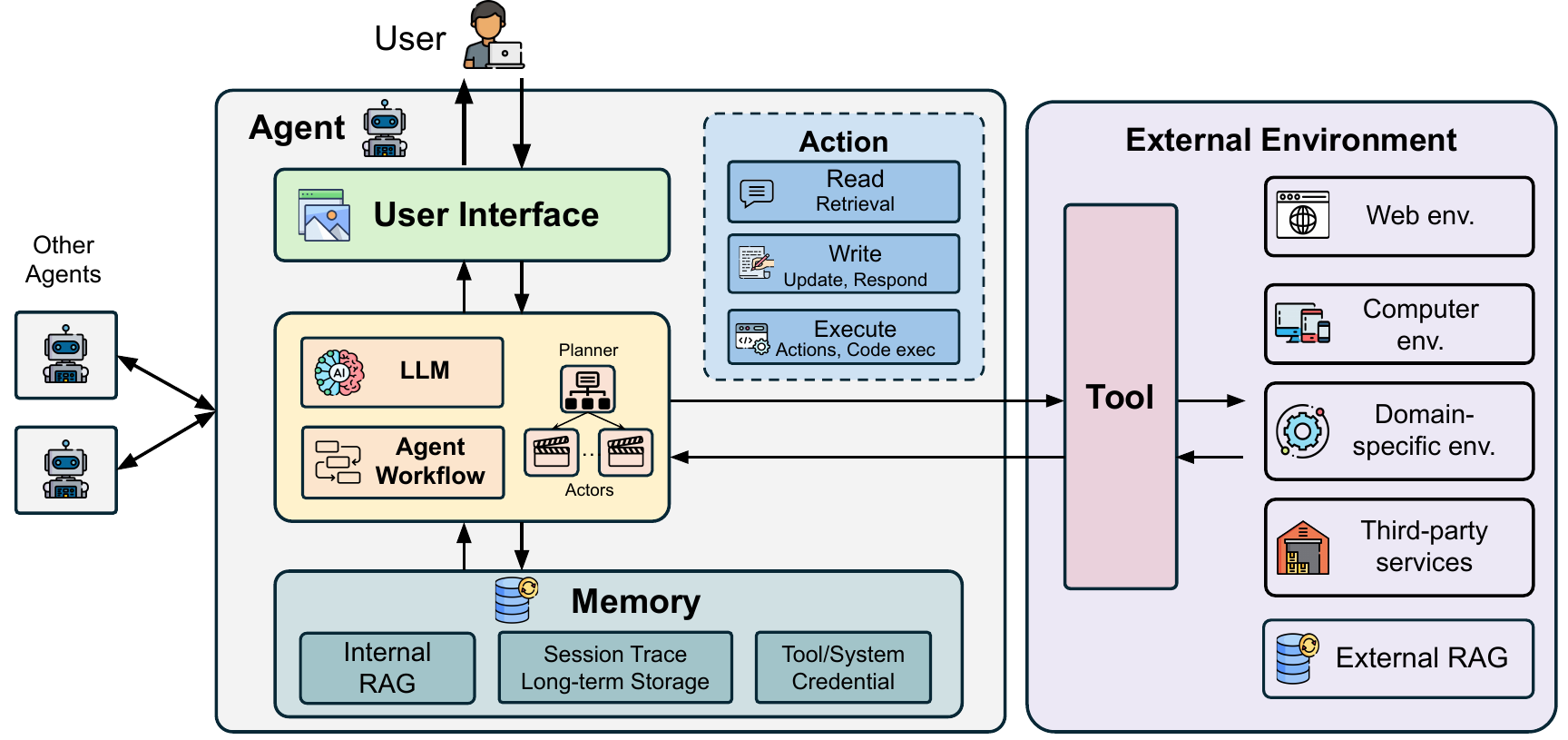}
    \caption{Overview of an AI Agent's structure.}
    \Description{A diagram illustrating the structural components of an AI agent, including the LLM model, memory, tool use, and planning modules, and how they interact with each other and with external environments.}
    \label{fig:agent}
\end{figure}

\PP{Tools} Tools are functions that an agent uses to interact with its external environment.
As shown in~\autoref{fig:agent}, there are typically retrieval tools and execution tools based on agent's actions.
Here, retrieval tools are used to collect information from the external environment (e.g., search or read file), and execution tools are used to make changes to the environment (e.g., write files, send emails, or run a command).
Tools can be implemented through standardized protocols (e.g., MCP~\cite{mcp}).
It is worth noting that the tools can be developed by the agent designer or third-party providers~\cite{mcp-server,gorilla}.

\PP{External environment}
An external environment is the external context in which an agent interacts to accomplish tasks.
Different agents operate in different environments, and the environment largely determines the agent's attack surface and defense priorities.
\emph{Web agents} interact with browsers and process arbitrary untrusted content from diverse web sources, making them particularly susceptible to indirect prompt injection and data exfiltration.
\emph{Coding agents} operate within IDEs and local file systems, where the primary risks involve unauthorized code execution and workspace manipulation.
\emph{Computer use agents} control GUI elements across applications via accessibility APIs or screen-based interfaces, exposing them to UI-based injection and broad system-level access.
The agent interacts with its environment (e.g., access, read, and write to a SQL database) through tools.



\PP{Overall structure}
An AI agent receives a user query and automatically takes a sequence of actions to assist the user.
The agent workflow orchestrates this process by coordinating LLM components that typically serve two roles. \emph{Planners} decompose user tasks into step-by-step plans, and \emph{actors} execute individual steps by invoking tools, generating content, or querying memory.
In simple agents, a single LLM fulfills both roles. In more complex systems, separate LLMs are dedicated to planning and execution, enabling modular control and finer-grained security boundaries.
For example, when asked to update a file, the planner determines the required steps, and the actor sequentially calls a file-read tool, generates new content, and calls a file-write tool to perform the update.
This architecture naturally extends to \emph{multi-agent systems} (MAS), where multiple specialized agents collaborate to accomplish complex tasks.

\PP{Agentic Systems vs. Traditional Systems}
The key uniqueness of agentic systems compared to traditional systems arises in the following aspects.
First, agentic systems combine traditional programs with AI model reasoning, whereas traditional systems rely mainly on symbolic logic.
This hybrid architecture makes agentic systems more flexible and adaptive.
Second, workflows in traditional software are largely pre-programmed, whereas agentic systems can dynamically decide their workflows and actions based on different tasks and~\emph{input contexts}.
Third, agentic systems use vectorized memory that supports semantic-based retrieval.
In contrast, traditional systems use structured memory access through predefined queries.

\begin{table*}[t]
\centering
\footnotesize
\caption{Agent design dimensions and corresponding flexibility for each dimension.
The agent design dimensions are conceptually orthogonal, but it practice, their functional implementation can introduce dependencies between them.}
\label{tab:design-dimensions}
\begin{tabularx}{\linewidth}{@{}p{1.6cm}p{3.5cm}p{3.5cm}X@{}}
\toprule
\textbf{Dimension} & \textbf{Level 1}: Least flexible & \textbf{Level 2}: Moderately flexible & \textbf{Level 3}: Most flexible \\
\midrule
\textbf{\dimtrust}
& No external data~\cite{achiam2023gpt,touvron2023llama1,team2023gemini}
& Predefined external data~\cite{lewis2020retrieval}
& Arbitrary external data~\cite{autogpt,gorilla,mcp-server} \\
\makecell[l]{\textbf{Access}\\\textbf{Sensitivity}}
& No sensitive data~\cite{achiam2023gpt,touvron2023llama1,team2023gemini}
& Predefined sensitive data~\cite{chatgpt-connectors}
& Arbitrary sensitive data~\cite{cursor,comet-browser} \\
\textbf{\dimworkflow}
& Simple chatbot~\cite{achiam2023gpt,touvron2023llama1,team2023gemini}
& Fixed, Developer-defined~\cite{workflow-agent}
& Dynamic, LLM-defined~\cite{wei2022chain,yao2023react,autogpt} \\
\textbf{\dimaction} & LLM response only~\cite{achiam2023gpt,touvron2023llama1,team2023gemini}
& LLM response, Retrieval~\cite{lewis2020retrieval,chatgpt-connectors}
& LLM response, Retrieval, Execution~\cite{cursor,comet-browser} \\
\textbf{\dimmemory}
& No memory~\cite{achiam2023gpt,touvron2023llama1,team2023gemini}
& Transient session memory~\cite{langchain-chat-history}
& Persistent memory across sessions~\cite{openai-memory,lee2023prompted} \\
\textbf{\dimtool}
& No tool~\cite{achiam2023gpt,touvron2023llama1,team2023gemini}
& Known tools~\cite{toolformer}
& Arbitrary tools~\cite{gorilla} \\
\makecell[l]{\textbf{User}\\\textbf{Interface}}
& Text-only~\cite{achiam2023gpt,touvron2023llama1,team2023gemini}
& Web-based image preview~\cite{nanobanana}
& Interactable Web elements~\cite{gemini-canvas,chatgpt-shared-links} \\
\bottomrule
\end{tabularx}
\end{table*}

\subsection{Design Dimensions and Security Implications}
\label{subsec:agent_design}

Based on the agent structure in~\autoref{fig:agent}, we further identify seven agent design dimensions, each of which represents a continuous spectrum of flexibility.
In~\autoref{tab:design-dimensions}, we illustrate three representative levels (least, moderately, and most flexible) for simplicity, but agents can operate at any point along this spectrum.

\PP{Input Trust}
This categorizes the trustworthiness of external data sources on which an agent is relying. It captures the progression from using no external data (highest trust) to relying on arbitrary, potentially untrusted external data sources (lowest trust but highest flexibility). As agents access more diverse and potentially untrusted information sources, they gain improved knowledge and decision-making capabilities, but face increased security risks from compromised or malicious data sources.

\PP{Workflow}
This refers to the action sequences of the agent and determines who defines these sequences, analogous to the code in traditional software. Flexibility increases as it shifts from no complex workflows (simple chatbots) to developer-defined workflows, and ultimately to LLM-defined dynamic workflows. This progression allows agents to adapt their execution patterns from rigid, predetermined sequences to flexible, context-aware task planning.

\PP{Access Sensitivity}
This categorizes an agent's level of access to sensitive data within the system or environment. Flexibility increases as agents gain access from no sensitive data to known sensitive data sources, and finally to arbitrary sensitive data. Higher levels enable agents to handle more complex tasks that require sensitive information access while significantly expanding both their operational scope and the potential impact of security breaches.
In particular, agents that handle personally identifiable information (PII), such as names, email addresses, financial records, or medical data, face elevated risks of data leakage through unintended tool outputs, logging, or exfiltration attacks.


\PP{Action}
Action describes what operations an agent can perform beyond text generation.
Flexibility increases from response-only, to read-only actions such as retrieval or querying, and eventually to environment-modifying operations including file edits, command execution, or API calls.
As actions become more powerful, both task completion and potential security risks grow.

\PP{Tool}
Tool defines the scope of external tools an agent can invoke.
The spectrum begins with no tool usage, expands to a fixed set of curated tools, and culminates in the ability to use or even select arbitrary tools.
Broader tool access enhances functionality but expands the attack surface substantially.

\PP{Memory}
Memory characterizes how an agent stores and retrieves information over time.
Agent ranges from memory-less operation, to transient session-level memory, and finally to persistent memory spanning multiple sessions.
Increasing reliance on memory supports personalization and long-term reasoning while introducing risks such as memory poisoning and private data leakage.

\PP{User Interface}
This dimension defines how users interact with an agent, with flexibility ranging from text-only interactions to graphical interactions through image preview or simple web interfaces, and ultimately to multi-modal interactions such as web, terminal, and integrated development environments (IDEs).
Richer interfaces enable more expressive workflows and interactions, but also broaden the range of possible attack vectors.

\PP{Security Implication}
In general, there is a trade-off between flexibility and security.
More specifically, a more flexible agent architecture broadens the attack surface and enables more diverse attack vectors.
For example, the attack targets for a simple chatbot are the input data and the model parameters; the sole attack vector runs from user input to model output.
In a multi-agent system, however, adversaries can target LLMs, shared memory stores, tools, and the external environment.
Attack vectors exist among system components (LLM to tools, memory to LLMs).
First, more complex memory designs and user interfaces further enable additional attack opportunities.
Second, as the agent’s input space becomes more flexible, attackers can more easily inject malicious data or instructions, facilitating poisoning and injection attacks.
Third, greater workflow flexibility increases the likelihood of control-flow hijacking, allowing attackers to redirect agent execution and launch various attacks.

\begin{figure*}[t]
    \centering
    \includegraphics[width=.9\textwidth]{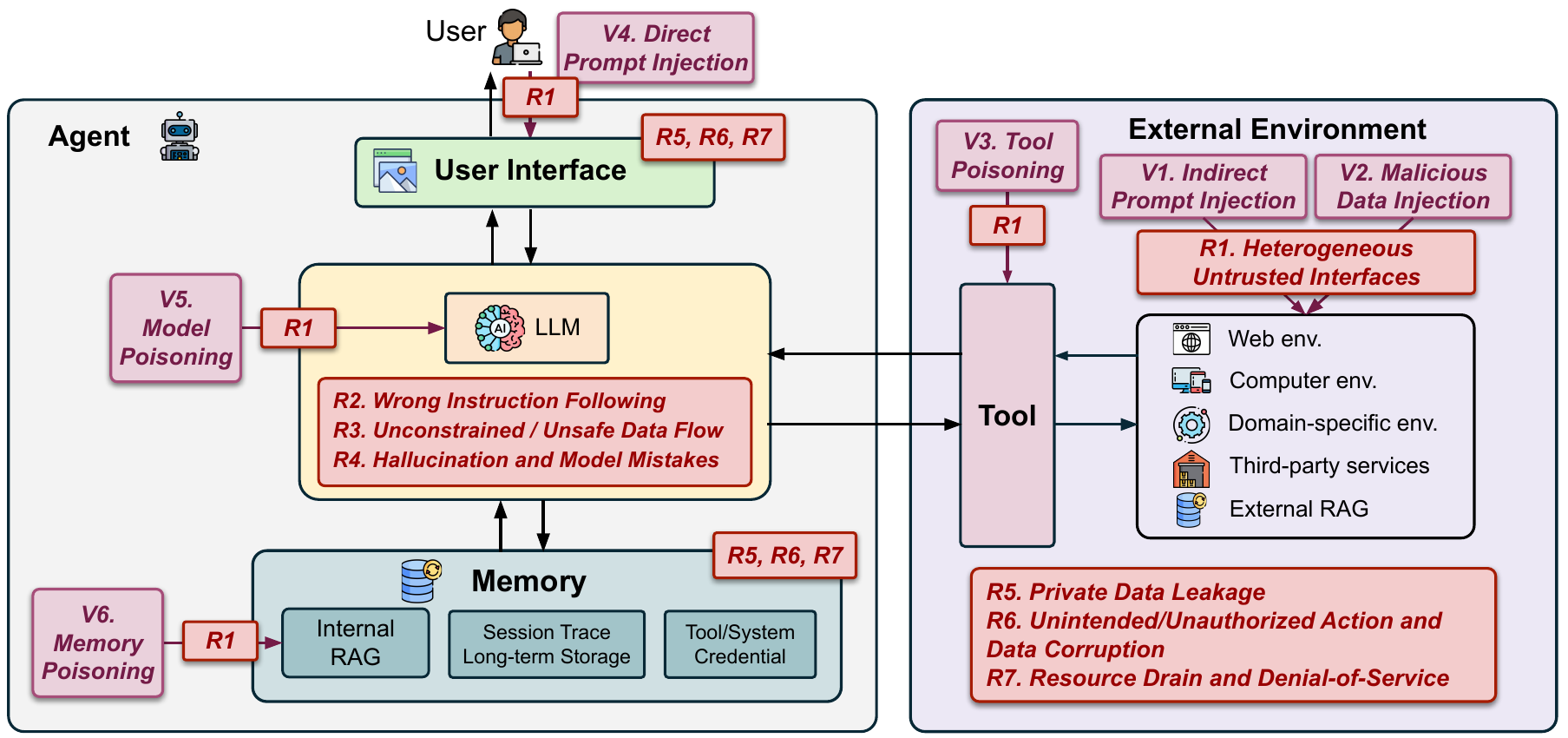}
    \caption{Demonstrations of attack vectors~(\vectorindirect-\vectormemory) and security risks (\riskattack-\riskavailability) against AI agents.}
    \Description{A diagram mapping six attack vectors (V1-V6) to seven security risks (R1-R7) in AI agent systems, organized by threat models including prompt injection, data injection, tool misuse, and memory poisoning.}
    \label{fig:attacks}
\end{figure*}

\section{Attack Landscape of Agentic AI Systems}
\label{sec:attack}

In this section, we conduct a comprehensive analysis of attacks against agentic systems by identifying key attack vectors~(\autoref{subsec:attack_threat}) and risk taxonomy~(\autoref{subsec:risk}), and by examining how design dimensions correspond to these risks and how different risks interact.
\autoref{fig:attacks} provides an overview of the attack landscape.

\subsection{Attack Vectors}
\label{subsec:attack_threat}


We discuss attack vectors in AI agents under three threat models, classified based on attackers' access to the agent at the time of attack execution rather than during the attack's development.
Assumptions specific to the attack development are considered when we describe attack methods (\autoref{subsec:attack_methods}). 


\PP{External Adversary}
An attacker is in an external environment and cannot directly interact with an agent, but the attacker can manipulate external resources that the agent may retrieve and process.
This threat model represents the most constrained yet highly realistic scenario.

\noindent\underline{V1. Indirect prompt injection}\label{vector:indirectinjection}~\cite{greshake2023not,github-mcp-exploited,lee2024prompt,liao2024eia,agentvigil,wu2024agentattack}: attackers inject malicious instructions into the external environment that the agent interacts with, such as public web pages or documents.
When the agent retrieves content from this environment, it may also retrieve and execute the malicious instructions.

\noindent\underline{V2. Malicious data injection}\label{vector:datainjection}~\cite{spracklen2025we,patlan2025real}:
The attacker can inject malicious non-prompt data.
When such data is consumed during sensitive operations, it can trigger security failures and breaches.
Examples include malicious software packages~\cite{spracklen2025we} or attacker-supplied values for sensitive financial parameters~\cite{patlan2025real}.

\noindent\underline{V3. Tool poisoning and manipulation}\label{vector:tool}~\cite{jumping-the-line,shi2025prompt}:
Attackers inject malicious instructions into the names or descriptions of external tools that agents interact with.
They can also inject malicious payloads into tool implementations. 

\PP{User-level Adversary}
Here, attackers have access to the agent inputs and can directly feed malicious contents to the agent~\cite{liu2023prompt} or inject them into otherwise benign user inputs~\cite{imprompter-attack,fu2023misusing}.

\noindent\underline{V4. Direct prompt injection}\label{vector:directinjection}~\cite{liu2023prompt,imprompter-attack,fu2023misusing}:
Attackers can control parts of otherwise benign inputs and append malicious instructions to user inputs.

\PP{Internal Adversary}
Attackers can access some or all components inside the agent, which represents the strongest assumption.
This attack vector poses the most severe threat, as attackers can control the agent's internals, but it is less practical.

\noindent\underline{V5. Model poisoning}\label{vector:model}~\cite{yang2024watch,wang2024badagent}:
The attacker injects a backdoor into the LLM, which can be activated during inference to enable malicious behavior.

\noindent\underline{V6. Memory poisoning}\label{vector:memory}~\cite{agentpoison,zou2024poisonedrag,dong2025practical,zhong2023poisoning}:
An attacker can directly manipulate the agent's memory to inject malicious instructions or false knowledge. 
Alternatively, the attacker can leak sensitive user data from the memory.

\subsection{Security Risks}
\label{subsec:risk}

We present a comprehensive taxonomy of security risks in agents, categorized by components that can be targeted by different attack vectors. 
This taxonomy provides a systematic map of the entire security risk landscape for agentic systems.

\PP{R1. Heterogeneous untrusted interfaces}\label{risk:attack-surface}
Compared to standalone LLMs, agentic systems expose multiple heterogeneous interfaces to users, including external data sources that agents retrieve and process, persistent memory stores that accumulate over time, and third-party tools with various trust levels.
These interfaces can be leveraged by attackers as attack 
entry points, which introduce way larger attack surfaces compared to standalone LLMs.

\PP{R2. Wrong instruction following}\label{risk:wrong-instruction}
The agent follows malicious prompts injected by attackers rather than the intended instructions from benign users or system developers. 
This occurs because inputs from external or attacker-controlled sources are processed by the model inside the agent and can therefore influence its behavior.

\PP{R3. Unconstrained/unsafe data flow}\label{risk:data-flow}
LLMs suffer from unconstrained data flow due to their stochastic nature.  
In agentic systems, this manifests as data flowing freely from any input from untrusted interfaces~(\riskattack) to any output (e.g., user responses or subsequent tool calls), unlike traditional systems where data propagation is regulated by programming languages, type systems, and access controls.

This leads to consequential risks: private data leakage~(\riskconfidentiality), data corruption~(\riskintegrity), and resource drain~(\riskavailability).
For example, when an agent retrieves untrusted web content and produces a URL in its response, that URL may contain attacker-controlled data. Automatically fetching such URLs enables data exfiltration~\cite{cve-2025-32711,github-data-exfiltration}, resulting in~\riskconfidentiality.

\PP{R4. Hallucinations and model mistakes}\label{risk:hallucination}
Models often hallucinate and generate incorrect information, and this problem becomes more severe in agent-environment interactions because agents act on hallucinated content, creating real-world consequences beyond misinformation, such as accessing attacker-controlled resources. 
Attackers exploit this behavior through package hallucination attacks~\cite{spracklen2025we}, where they register malicious packages with names that LLMs frequently hallucinate. 
This turns a model limitation into a reliable attack vector for code injection and supply chain compromise.

\PP{R5. Private data leakage}\label{risk:confidentiality}
Agents handle sensitive data across multiple components such as user conversations, persistent memory, tool credentials, and environment resources, creating opportunities for unauthorized data access. 
Attackers exploit interactions among these components to exfiltrate private information. 
Examples include manipulating agents to send data to attacker-controlled servers, triggering automatic URL fetches that leak data through request parameters, or abusing domain-specific vulnerabilities such as SSRF, XSS, path traversal, or SQL injection to reach protected resources. 
Recent incidents~\cite{march-20-chatgpt,mozilla-meta-ai,langsmith-leak} show real confidentiality violations where agent vulnerabilities exposed users' chat histories and personal information. 

\PP{R6. Unintended/unauthorized action and data corruption}\label{risk:integrity}
Agents can violate integrity through two closely related risks, unintended and unauthorized actions, and data corruption, both involving unauthorized changes to internal or external state.
Unintended actions make irreversible state changes (e.g., unauthorized purchases, arbitrary code execution), while data corruption directly modifies stored resources (e.g., corrupting files or databases).
At the user interface, agents may provide false or misleading information.
At memory components, agents may inject poisoned knowledge or malicious instructions~\cite{agentpoison} that corrupt subsequent behavior.
At the environment level, agents can be manipulated to execute unauthorized modifications through command injection, SQL injection, or file system manipulation.

\PP{R7. Resource drain and denial-of-service}\label{risk:availability}
Agents introduce availability risks through their consumption of computational resources, API calls, and interactions with external systems~\cite{kumar2025overthink}.
Attackers exploit this autonomy to trigger costly API calls, force infinite execution loops, or cause excessive memory use, creating denial-of-service conditions that can render the agent and external systems unusable or economically unsustainable.

\PP{Other risks in agentic systems}
Attackers can also target the agentic AI system itself, rather than the user or surrounding environment, by stealing system prompts, tool descriptions, and configurations that encode proprietary knowledge~\cite{shen2024prompt,yang2025prsa}, undermining agent developers and tool providers.
Agentic systems may also expose side-channel risks through observable behaviors such as tool execution timing, API call sequences, or network traffic patterns.
To the best of our knowledge, we are not aware of prior work studying such side-channel attacks in agentic systems, and thus defer deeper analysis to future work.

\PP{From Attack Vectors To Risks}
These risks can be exploited via various attack vectors under different threat models. The wrong instruction following risk (\riskwronginst) can be triggered by all attack vectors and carried out in all threat models. Private data leakage (\riskconfidentiality) can result from indirect prompt injection (\vectorindirect), data injection (\vectordata), or memory poisoning (\vectormemory). Unauthorized actions and data corruption (\riskintegrity) are commonly executed through indirect prompt injection (\vectorindirect) but can also be exploited through direct injection (\vectordirect), tool poisoning (\vectortool), or model poisoning (\vectormodel). Resource drain (\riskavailability) can be triggered through any external or user-level vector.

\begin{figure*}[t]
    \centering
    \includegraphics[width=0.9\textwidth]{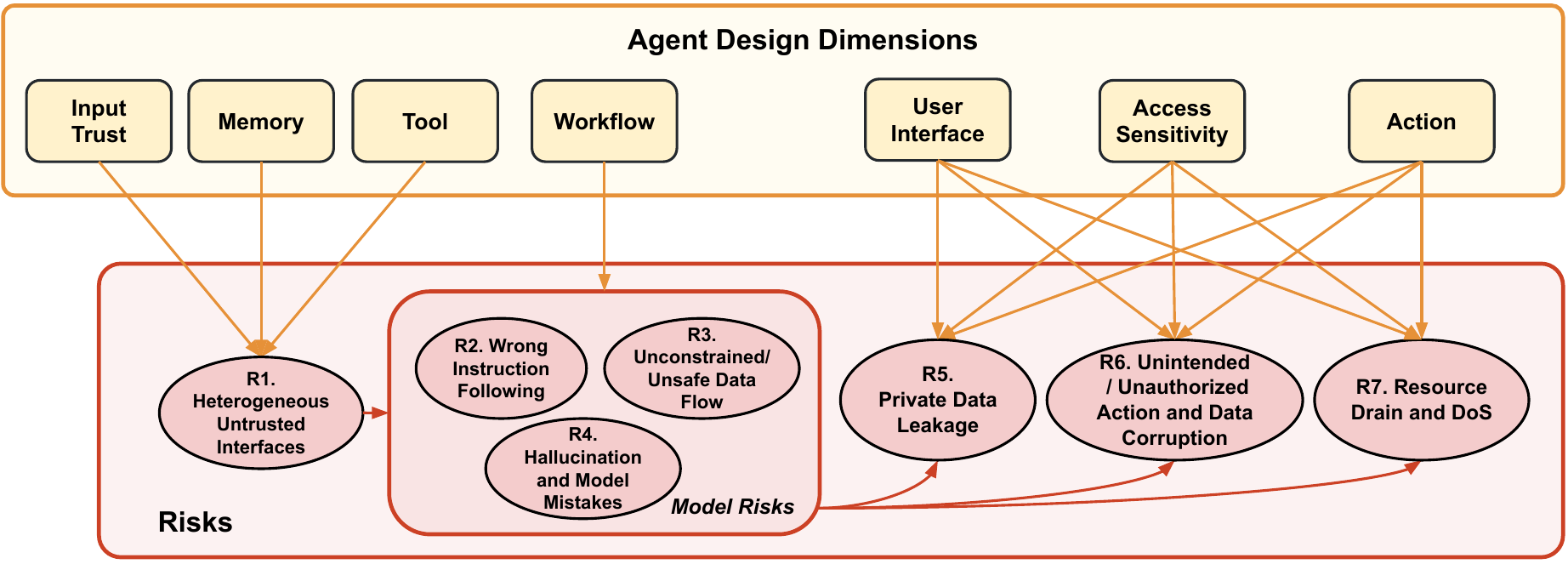}
    \caption{Relationship between agent design dimensions and risks, and the connection across different risks.}
    \Description{A matrix or graph showing how each agent design dimension (trust, sensitivity, workflow, action, memory, tool, interface) maps to and influences specific security risks R1 through R7, with connecting arrows indicating dependencies between risks.}
    \label{fig:design-risk}
\end{figure*}

\subsection{System-level Analysis of Agent Risks}
\label{subsec:system_analysis}
\autoref{fig:design-risk} illustrates how agent design dimensions map to security risks and how these risks interact to amplify system-level threats.

\PP{Design Dimensions to Risks}
Agent design dimensions map to distinct risk categories, and we observe that greater flexibility amplifies security risks~(\autoref{fig:design-risk}).
The \dimtrust, \dimmemory, and \dimtool dimensions directly contribute to increased attack surface~(\riskattack).
Expanding external data sources enables indirect prompt injection, adding persistent memory creates memory poisoning targets, and incorporating third-party tools introduces supply chain risks.

The \dimworkflow dimension primarily drives model risks~(\riskwronginst, \riskunsafedataflow, \riskhallucination).
As workflows shift from simple chatbots to LLM-defined dynamic execution, flexible control flows provide more opportunities for hijacking agent reasoning, worsening wrong instruction following and unconstrained data flow.
Hallucination risks amplify when agents dynamically select tools and resources, as incorrect model outputs directly trigger real-world actions.

The \dimsensitivity, \dimaction, and \diminterface dimensions determine the severity of consequence risks~(\riskconfidentiality, \riskintegrity, \riskavailability).
Granting agents access to more sensitive data amplifies the impact of violations, as compromised agents can leak, corrupt, or drain more valuable resources.
Expanding action capabilities from response-only to execution transforms information disclosure into environment corruption.
Complex user interfaces introduce new attack channels through automatic URL fetching (confidentiality), executing destructive commands (integrity), and interface manipulation (availability).

\PP{Risk Interactions and Amplification}
Risks in agentic systems~(\riskattack-\riskavailability) interact in a cascading manner, where initial failures propagate across components and amplify system-level threats.
An expanded attack surface~(\riskattack) increases entry points for attacker-controlled data, allowing malicious inputs to reach and exploit model risks~(\riskwronginst, \riskunsafedataflow, \riskhallucination).
For example, indirect prompt injection through external data can trigger wrong instruction following, which then redirects agent behavior.
Model risks then amplify consequence risks~(\riskconfidentiality, \riskintegrity, \riskavailability).
Wrong instruction following can lead to data exfiltration (confidentiality), data corruption (integrity), or excessive resource consumption (availability).
Unconstrained data flow heightens confidentiality and integrity risks by enabling data leakage and malicious code execution.
Hallucination increases integrity risks by causing agents to operate on fabricated resources, leading to data leakage and corruption.
As a concrete example, EchoLeak~\cite{aim-echoleak,cve-2025-32711} demonstrates how a malicious document embedded in an enterprise email exploits heterogeneous untrusted interfaces~(\riskattack), triggers unconstrained data flow through the agent's retrieval pipeline~(\riskunsafedataflow), and ultimately exfiltrates sensitive user data to an attacker-controlled server~(\riskconfidentiality), all without any user interaction.

\subsection{Attack Methods}
\label{subsec:attack_methods}

Attack methods refer to the techniques that construct attack paths and generate attack payloads.
Due to the attack complexity, most existing attacks heavily rely on human efforts to construct attack paths and payloads~\cite{cve-2024-5565,cve-2025-32711}.
For example, the injection points for just-in-time injection and memory poisoning attacks are almost always set manually.
The attack payloads (i.e., malicious instructions) for prompt injection attacks are mainly generated based on pre-specified attack patterns, such as role-playing scenarios~\cite{debenedetti2024agentdojo}, delimiter confusion using special characters~\cite{delimiters_url, pi_against_gpt3}, or instruction reset commands that attempt to ask LLMs to ignore previous context~\cite{perez2022ignore, schulhoff2023ignore}.

Recent research has started to explore automated methods for generating attack payloads and injection points.
For attack payload, prompt injection attacks design specific fuzzing approaches for AI agents~\cite{agentvigil,yu2025promptfuzz}, as well as training small attack models for malicious instruction generation~\cite{zou2023universal,wu2024agentattack,liao2024eia}. 
Memory poisoning techniques employ several approaches focused on maximizing retrieval likelihood while maintaining content credibility.
Semantic injection represents the most systematically studied approach, crafting factually incorrect content with high semantic similarity to target queries through embedding optimization techniques.
This method leverages contrastive learning principles to position malicious content close to legitimate queries in the embedding space, ensuring preferential retrieval by vector similarity search mechanisms~\cite{zou2024poisonedrag}.
Advanced variants employ gradient-based optimization to craft adversarial passages that achieve optimal embedding similarity while maintaining semantic coherence~\cite{universal-prompt,zhong2023poisoning,zhang2024goal}.
In general, the community can benefit from more automated end-to-end attack/red-teaming methods for agents, which can be used as in-house testing tools by agent developers.

\begin{figure*}
    \centering
    \includegraphics[width=\linewidth]{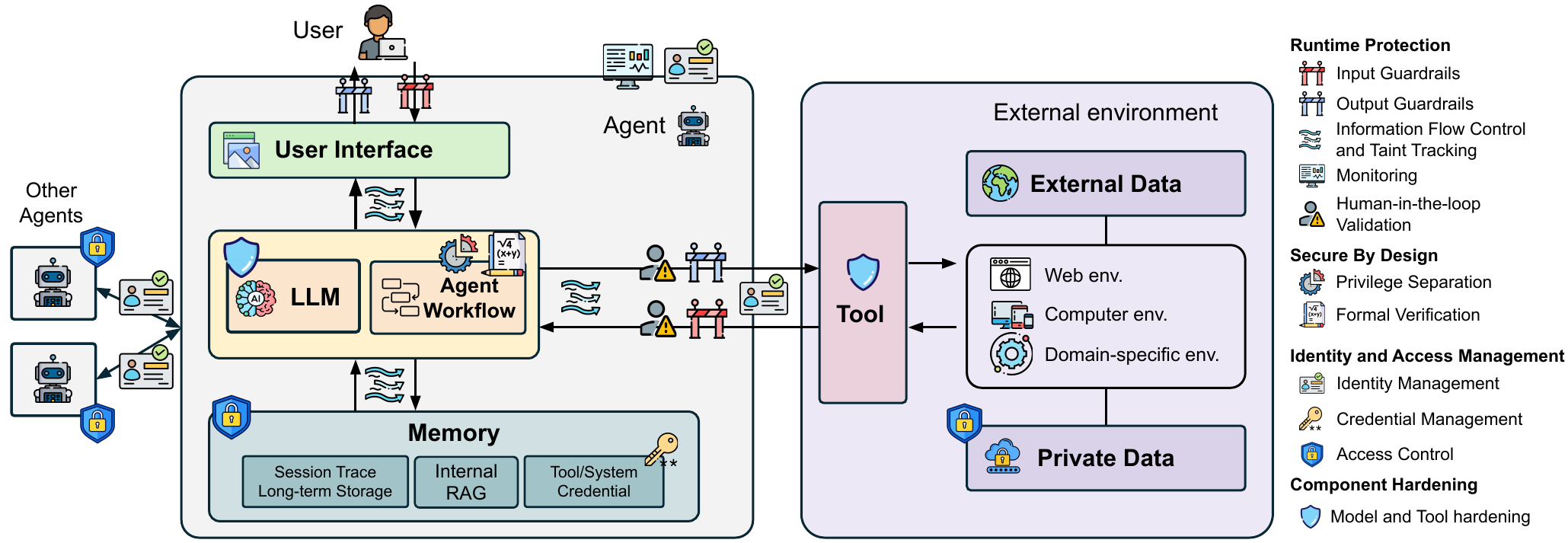}
    \caption{The defense landscape of AI agents. We illustrate key defense mechanisms and where they can be applied within the agent system.}
    \Description{A layered diagram of an AI agent system annotated with defense mechanisms at each layer, including input guardrails, output guardrails, access control, information flow control, monitoring, human-in-the-loop controls, privilege separation, and identity management.}
    \label{fig:defense}
\end{figure*}

\section{Defense Landscape of Agentic AI Systems}
\label{sec:defense}
The comprehensive defense landscape of AI agents has been largely underexplored.
We put together the defense landscape by examining security goals~(\autoref{sec:goals}) and identifying key defense mechanisms in different categories~(\autoref{sec:defense:runtime},\autoref{sec:defense:secure-by-design},\autoref{sec:defense:trust},\autoref{sec:defense:individual}).
For each category, we analyze their design dimensions and open challenges.
Finally, we discuss defense design principles~(\autoref{sec:defense:principles}).
\autoref{tab:defense-mechanisms} summarizes the defense mechanisms for AI agents and the risks that each defense covers, organized by category.
\autoref{fig:defense} illustrates the defense landscape.

\subsection{Security Goals}
\label{sec:goals}
In this section, we discuss the security goals for AI agents.
They are based on standard security principles, specifically the \emph{Confidentiality}, \emph{Integrity}, and \emph{Availability}~(CIA) triad that forms the foundation of traditional security frameworks.
Additionally, we introduce \emph{Contextual Security} as a new security goal for agentic systems.

\PP{Confidentiality}
Confidentiality ensures that all information is accessible only to authorized entities. 
It includes protecting system-level secrets (e.g., API keys, credentials), the agent's internal memory, users' private data, and LLM-related data (model parameters and system prompts).
Achieving this goal mitigates risks such as unsafe data flow~(\riskunsafedataflow), private data leakage~(\riskconfidentiality), and unconstrained data flows that expose credentials or sensitive information.

\PP{Integrity}
Integrity ensures that data and control flows within an agentic system and its external environment remain trustworthy and unaltered by unauthorized entities. 
It includes preventing tampering with the agent's memory, LLM outputs, tool results, and environmental data, as well as ensuring that agents are not manipulated into performing unintended actions by malicious instructions. 
Specific security goals associated with integrity can vary across agent types, depending on their components, data, and control flows.
Achieving this goal addresses unsafe data flow~(\riskunsafedataflow), hallucinations and model mistakes~(\riskhallucination), wrong instruction following~(\riskwronginst), and unintended/unauthorized actions and data corruption~(\riskintegrity).

\PP{Availability}
Availability protects the system from denial-of-service attacks and resource abuse~(\riskavailability), even when LLM inference or tools consume user resources.
This includes preventing token draining of LLM, as well as regulating the usage of resources in agent hosts and external system.

\PP{Contextual Security}
Recent studies define contextual security~\cite{tsai2025contextual,shi2025progent} as a critical security goal for AI agents.
Contextual security ensures that the agent's contexts are aligned with the agent's intended user tasks, preventing attacks aimed at manipulating contexts during agent execution.
It governs which context elements (e.g., system prompts, user goals, tool descriptions, retrieved snippets) are admissible and how they are prioritized to avoid instruction override, context drift, or unsafe tool selection, complementing confidentiality and integrity that protect data correctness and secrecy.

Contextual security stems from \emph{contextual integrity}~\cite{contextual-integrity}, a privacy framework that defines appropriate information flows based on social norms and context.
Contextual integrity has been applied to various software domains, including general programs~\cite{shvartzshnaider2019vaccine}, mobile applications~\cite{wijesekera2015android}, and IoT systems~\cite{jia2017contexlot}.
In the agentic setting, AirGapAgent~\cite{airgap} applies contextual integrity to ensure that agents only access information relevant to the current task context.
However, a recent position paper~\cite{shvartzshnaider2025position} finds that many works adopting contextual integrity fail to fully follow its principles, simplifying information flows and inadequately addressing the broader context and norms that contextual integrity requires.
In AI agents, open-ended inputs and diverse environments further complicate defining such contexts and norms.
While contextual integrity primarily addresses privacy concerns, contextual security extends this principle to security by ensuring that the agent's actions remain aligned with user intent and free from adversarial manipulation.

\begin{table*}[t]
\caption{Defense mechanisms for AI agents organized by category, as well as the risks each defense covers.}
\label{tab:defense-mechanisms}
\footnotesize
\begingroup
\begin{tabularx}{\textwidth}{p{0.12\textwidth}p{0.15\textwidth}X p{0.28\textwidth}}
\toprule
\textbf{Category} & \textbf{Mechanism} & \textbf{Representative works}
& \textbf{Covered risks} \\
\midrule
\multirow{7}{*}{\begin{tabular}[l]{@{}l@{}}\textbf{Runtime}\\\textbf{Protection}\end{tabular}}
& \textbf{Input guardrail} &
Prompt Guard2~\cite{llama-firewall}, PromptShield~\cite{jacob2025promptshield}, DataSentinel~\cite{liu2025datasentinel}, PromptArmor~\cite{shi2025promptarmor},
URL allowlist~\cite{google-mitigating-prompt}
&
\riskattack. Heterogeneous Untrusted Interfaces
\\
\cmidrule{2-4}
& \textbf{Output guardrail} &
CodeShield~\cite{codeshield}, Firewalled Agentic Networks~\cite{abdelnabi2025firewalls},
Task Shield~\cite{jia2024task}, GuardAgent~\cite{xiang2024guardagent}, Agrail~\cite{luo2025agrail},
AlignmentCheck~\cite{llama-firewall},
ControlValve~\cite{jha2025breaking},
Conseca~\cite{tsai2025contextual}, Progent~\cite{shi2025progent},
Maris~\cite{cui2025safeguard},
 &
\parbox[t]{0.3\textwidth}{\raggedright
\riskwronginst. Wrong Instruction Following\\
\riskhallucination. Hallucination and Model Mistakes\\
\riskconfidentiality. Private Data Leakage\\
\riskintegrity. Unintended/Unauthorized Action and Data Corruption}
\\
\cmidrule{2-4}
& \textbf{Information flow control and taint tracking} &
Permissive~\cite{siddiqui2024permissive}, MELON~\cite{zhu2025melon},
RTBAS~\cite{zhong2025rtbas}, AgentArmor~\cite{wang2025agentarmor},
PFI~\cite{pfi}, CaMeL~\cite{debenedetti2025defeating}, FIDES~\cite{costa2025securing}, ACE~\cite{li2025ace}, PrivacyChecker~\cite{wang2025privacy}
&
\parbox[t]{0.3\textwidth}{\raggedright
\riskunsafedataflow. Unconstrained/Unsafe Data Flow\\
\riskconfidentiality. Private Data Leakage\\
\riskintegrity. Unintended/Unauthorized Action and Data Corruption\\
}
\\
\cmidrule{2-4}
&
\textbf{Monitoring}
&
AgentAuditor~\cite{luo2025agentauditor}, AgentMonitor~\cite{naihin2023testing},
SentinelAgent~\cite{he2025sentinelagent}, Guardian~\cite{zhou2025guardian}
&
\multirow{2}{*}{\parbox[t]{0.3\textwidth}{\raggedright
\riskconfidentiality. Private Data Leakage\\
\riskintegrity. Unintended/Unauthorized Action and Data Corruption\\
\riskavailability. Resource Drain and Denial-of-Service
}}
\\
\cmidrule{2-3}
& \textbf{Human-in-the-loop} &
Wu et al.~\cite{wu2025towards}
&
\\
\addlinespace[10pt]
\midrule
\multirow{4}{*}{\begin{tabular}[l]{@{}l@{}}\textbf{Secure}\\\textbf{By Design}\end{tabular}}
& \textbf{Privilege separation} &
AirGapAgent~\cite{airgap}, f-secure~\cite{wu2024system}, CaMeL~\cite{debenedetti2025defeating}, FIDES~\cite{costa2025securing}, PFI~\cite{pfi}, IsolateGPT~\cite{wu2025isolategpt}, ACE~\cite{li2025ace}, IPIGuard~\cite{an2025ipiguard}, DRIFT~\cite{li2025drift}
&
\multirow{2}{*}{\parbox[t]{0.3\textwidth}{\raggedright
\riskwronginst. Wrong Instruction Following\\
\riskunsafedataflow. Unconstrained/Unsafe Data Flow
}}
\\
\cmidrule{2-3}
& \textbf{Formal verification} &
Formal-LLM~\cite{li2024formal}, VeriSafeAgent~\cite{lee2025safeguarding}, ShieldAgent~\cite{chen2025shieldagent}
&
\\
\midrule
\multirow{5}{*}{\begin{tabular}[l]{@{}l@{}}\textbf{Identity}\\\textbf{and Access}\\\textbf{Management}\end{tabular}}
& \textbf{Identity management} &
Authenticated delegation~\cite{south2025authenticated}, SAGA~\cite{syros2025saga}, Agent network protocol~\cite{chang2025agent}
& \multirow{2}{*}{\parbox[t]{0.3\textwidth}{\raggedright
\riskconfidentiality. Private Data Leakage\\
\riskintegrity. Unintended/Unauthorized Action and Data Corruption\\
\riskavailability. Resource Drain and Denial-of-Service
}}
\\
\cmidrule{2-3}
& \textbf{Access control} &
 ControlNet~\cite{yao2025controlnet}, Honeybee~\cite{zhong2025honeybee}, Bedrock~\cite{amazon-bedrock}, Authenticated delegation~\cite{south2025authenticated}, SAGA~\cite{syros2025saga}
&
\\
\cmidrule{2-4}
& \textbf{Credential management} &
Token vault~\cite{token-vault}
& \riskconfidentiality. Private Data Leakage
\\
\midrule
\multirow{4}{*}{\begin{tabular}[l]{@{}l@{}}\textbf{Component}\\\textbf{Hardening}\end{tabular}}
& \textbf{Model hardening} &
SecAlign~\cite{chen2025secalign}, StruQ~\cite{chen2024struq}, Instruction Hierarchy~\cite{wallace2024instruction}, InstructionalAgent~\cite{wu2024instructional}
& \parbox[t]{0.3\textwidth}{\raggedright
    \riskwronginst. Wrong Instruction Following\\
    \riskhallucination. Hallucination and Model Mistakes
}

\\
\cmidrule{2-4}
& \textbf{Tool hardening} &
Anthropic Connectors~\cite{anthropic-connectors}, ETDI~\cite{etdi}, MCP Context Protector~\cite{mcp-context-protector}, MCP Safety Audit~\cite{radosevich2025mcp}, MCIP~\cite{jing2025mcip}
&
\riskattack. Heterogeneous Untrusted Interfaces
\\
\bottomrule
\end{tabularx}
\endgroup
\end{table*}

\subsection{Runtime Protection}
\label{sec:defense:runtime}
Runtime protection mechanisms provide dynamic security enforcement during agent execution, detecting real-time threats and behaviors.


\subsubsection{Input Guardrail}
\label{sec:defense:input}

Input guardrails validate and sanitize possible input dimensions of agents, such as user input, tool retrieval results, and memory data.
They provide a \emph{first-line defense} against attack vectors by external and user-level adversaries, preventing malicious instructions and data from reaching their agent internals.
Input guardrails apply to both standalone models and agents.
For standalone models, input guardrails focus on detecting malicious inputs such as jailbreak prompts or harmful content~\cite{sharma2025constitutional,llama-firewall,jacob2025promptshield,liu2025datasentinel,shi2025promptarmor}.
These techniques remain effective in agentic settings to validate various inputs, especially those from external environments.
Specifically to agents, input guardrails address risks introduced by dynamic data retrieval and tool execution, verifying the trustworthiness of retrieved data. For instance, agents enforce URL allowlists~\cite{google-mitigating-prompt} to constrain data retrieval to trusted sources, mitigating risks from untrusted web content.

\PP{Design Dimensions}
Input guardrails can be characterized along three design dimensions: \emph{detection mechanism}, \emph{validation target}, and \emph{mitigation strategy}.
First, detection mechanisms face a fundamental tradeoff between security strictness and operational flexibility.
Rule-based detection offers strict but inflexible protection through predefined patterns~\cite{nemo-guardrails} or endpoint allowlists~\cite{google-mitigating-prompt, safe-browsing}. Model-based detection, on the other hand, trains small models~\cite{sharma2025constitutional,llama-firewall,liu2025datasentinel, jacob2025promptshield} or prompts LLMs~\cite{shi2025prompt} to detect or filter out malicious prompts. In general, LLMs are more generalizable and effective than small models in defense capabilities, but they also introduce more overhead and latency~\cite{wang2025sok}.
Second, validation target specifies what aspect of input data is being inspected. Content-based guardrails~\cite{sharma2025constitutional,llama-firewall,jacob2025promptshield,shi2025promptarmor} examine semantic input for malicious patterns or harmful content. In contrast, source-based guardrails~\cite{google-mitigating-prompt} verify data origin trustworthiness, uniquely addressing agents' dynamic retrieval from external sources like web search and databases.
Third, mitigation strategy defines how detected threats are handled. Guardrails can simply detect and filter out malicious input~\cite{sharma2025constitutional,llama-firewall,liu2025datasentinel,jacob2025promptshield}, sanitize the malicious portion before incorporating the input into the system~\cite{shi2025promptarmor}, or normalize input in a structured format with type system and validation~\cite{zod} to reduce the attack surface.
Alternatively, some model-based approaches neutralize threats without explicit detection by perturbing or smoothing inputs to remove adversarial perturbations~\cite{zhou2024robust,robey2025smoothllm}.

\PP{Limitations and Open Challenges}
As the input space becomes vast and diverse, it is challenging to establish universal security criteria.
Model-based detectors are often bypassed by adaptive attacks~\cite{andriushchenkojailbreaking,nasr2025attacker}, and rule-based detectors require extensive human effort and are difficult to generalize.
Consequently, input guardrails often suffer from false positives and false negatives, which negatively impact the agent's utility and security.
False positives are caused partly by the limited detection accuracy, but more importantly, by ambiguous definitions and boundaries between secure and insecure data and instructions.
False negatives allow attackers to bypass input guardrails and inject malicious inputs into the target agent.

\subsubsection{Output Guardrail}
\label{sec:defense:output}

Output guardrails perform security checks on outbound results of agents, such as responses to users and tool invocations that interact with external systems.
They can complement input guardrails by preventing attacks that appear benign from input prompts but result in malicious actions~\cite{cve-2025-32711}.
Output guardrails for LLMs typically focus on detecting harmful outputs, using classifiers or programmable rules~\cite{sharma2025constitutional,nemo-guardrails,codeshield,heimdallm}.
Agent-specific output guardrails validate tool usage and action sequences~\cite{xiang2024guardagent,chen2025shieldagent,shi2025progent,tsai2025contextual,luo2025agrail,jia2024task}, ensuring context-dependent policies and alignment with the user intent.
In multi-agent settings, output guardrails can also enforce inter-agent communication policies~\cite{abdelnabi2025firewalls} and permitted control-flow graphs that prevent unauthorized agent transitions~\cite{jha2025breaking}.

\PP{Design Dimensions}
Similar to input guardrails, output guardrails can be characterized along two design dimensions: \emph{detection goal} and \emph{detection mechanism}.

First, output guardrails detect harmful content in LLM outputs~\cite{sharma2025constitutional,nemo-guardrails}, unsafe code~\cite{codeshield,heimdallm}, and unsafe actions in tool usage~\cite{shi2025progent,tsai2025contextual}. Further, output guardrails perform alignment checks to ensure agent's behavior remain within user intent~\cite{llama-firewall,jia2024task}, data privacy protection~\cite{cui2025safeguard}, domain-specific policy enforcement~\cite{xiang2024guardagent,chen2025shieldagent,luo2025agrail}, and runtime-aware contextual security enforcement~\cite{tsai2025contextual,shi2025progent}.

Second, mechanisms of output guardrails range from rule-based pattern matching~\cite{codeshield,heimdallm} to model-based classifiers~\cite{sharma2025constitutional,nemo-guardrails,llama-firewall}.
Hybrid approaches adopt a structured policy framework assisted by models to process unstructured data, enabling flexible security enforcement~\cite{shi2025progent,tsai2025contextual,xiang2024guardagent,chen2025shieldagent,luo2025agrail,cui2025safeguard}.

\PP{Limitations and Open Challenges}
Output guardrails share similar limitations with input guardrails, including false positives and negatives due to unclear security criteria, and a trade-off between rule-based and model-based solutions.
Compared to input guardrails, output guardrails consume more computational resources and time, as they have a dependency on the LLM's output and need to process more data~\cite{wang2025sok}.
For guardrail methods, it is critical to balance the trade-off between security and utility by minimizing the latency and reducing guardrail false positives.

\subsubsection{Information Flow Control and Taint Tracking}
\label{sec:defense:flow}

Information Flow Control~(IFC)~\cite{myers1999jflow,denning1976lattice,bell1973secure,biba1977integrity} and taint tracking~\cite{newsome2005dynamic} restrict the data flow of information within a system.
At a high level, such methods assign each data a security label from a predefined information flow lattice~\cite{denning1976lattice} and propagate it along the agent execution, detecting unsafe information flows that violate the lattice constraints.

\PP{Design Dimensions}
IFC and taint tracking designs vary across \emph{security goals} and \emph{mechanisms}, spanning non-agentic LLM outputs and agent tool executions.

First, previous works provide integrity and confidentiality protection in agents.
Integrity protection blocks untrusted inputs from influencing tool call decisions~\cite{zhu2025melon,zhong2025rtbas,pfi}.
Confidentiality protection prevents sensitive data from reaching untrusted sinks~\cite{debenedetti2025defeating,costa2025securing,li2025ace,wang2025privacy}.

Second, IFC and taint tracking mechanisms include symbolic variable-based, multi-execution-based, and model-based approaches.
Multi-execution~\cite{siddiqui2024permissive,zhu2025melon} measures the influence of an input on an output by performing the LLM inference multiple times with and without the input.
Variable-based approaches replace data with trackable variables to mitigate over-tainting while retaining deterministic information flow guarantee~\cite{pfi,debenedetti2025defeating,costa2025securing,li2025ace}.
Model-based approaches request LLMs to inspect information flow given agent traces~\cite{wang2025agentarmor,li2025safeflow,zhong2025rtbas,wang2025privacy}.

\PP{Limitations and Open Challenges}
Existing IFC and taint methods incur substantial runtime overhead with multi-execution or variable-based reasoning, limiting practicality for latency-sensitive agents.
They may also suffer from label creep, where conservative propagation renders agents unusable unless automated and safe declassification rules are devised.
Bridging these gaps requires lightweight information flow tracking and principled policies for relaxing security labels when it is safe to do so, without compromising security.

\subsubsection{Monitoring}
\label{sec:defense:monitor}
Under the dynamic and unpredictable nature of AI agents, monitoring offers system-wide visibility by checking inputs, outputs, and intermediate states.
This holistic view helps surface distributed threats where no single input or action appears malicious in isolation but collectively constitutes a malicious goal~\cite{wen2025adaptive,yueh2025monitoring}.
Monitoring can be important to observe interactions across tools and services over long runs, especially for multi-agent systems.

\PP{Design Dimensions}
Monitoring design varies across \emph{detection goals} and \emph{log granularity}.

First, monitoring systems target different threat categories, including anomaly detection over long agent trajectories~\cite{naihin2023testing,luo2025agentauditor,he2025sentinelagent}, and interaction-graph monitoring in multi-agent systems~\cite{zhou2025guardian}.

Second, agent activity log ranges from coarse summaries of actions and tool calls to fine-grained traces that capture intermediate reasoning steps and parameters. Finer granularity improves detection power but increases storage, computation, and privacy exposure~\cite{chan2024visibility}.

\PP{Limitations and Open Challenges}
While monitoring techniques provide a holistic view of agent execution, they suffer from fundamental limitations of runtime defenses, such as inaccuracy issues and performance overhead.
Agent behaviors are stochastic and context-dependent, making it hard to distinguish benign actions from harmful ones.
Static rules miss novel attacks, whereas adaptive models incur overhead and remain susceptible to evasion.
Moreover, long-lived executions would further accumulate logs, for which storage overhead and privacy controls remain largely unexplored.

\subsubsection{Human-In-The-Loop Validation}
\label{sec:defense:human}

Traditional security employs user consent mechanisms for app installation~\cite{felt2012android} and sensitive data access~\cite{ios-permission}.
In agentic settings, human-in-the-loop validation allows users to validate agent behavior and tool usage, providing user-customized control over security decisions.
While limited literature has been dedicated to discussing human-in-the-loop validation for agents, contemporary coding agents such as GitHub Copilot~\cite{github-copilot}, Gemini Code Assist~\cite{gemini-code-assist}, Cursor~\cite{cursor}, and Codex~\cite{codex} ask for user approval when writing a file or executing command-line commands.
Agent defense systems~\cite{pfi,debenedetti2025defeating,costa2025securing,wu2025isolategpt} often leverage human-in-the-loop validation when agents attempt actions that violate defense policies.

\PP{Design Dimensions}
Human-in-the-loop has three design dimensions: \emph{validation scope}, \emph{user alert}, and \emph{recurrence policy}.

First, the \emph{validation scope} defines the scope of actions for which the agent requests user approval.
Existing coding agents typically require approval for terminal commands, file accesses outside the current workspace, or destructive actions like file deletion.

Second, the \emph{user alert} provides context to support user approval decisions. Effective alerts should clearly explain the agent's intended action, associated risks, and potential consequences to support informed decision-making.

Third, the \emph{recurrence policy} determines how often alerts are presented. To reduce approval frequency and decision fatigue, the agent can offer an option to remember the user's choice. Similar to mobile permission systems~\cite{felt2012android}, these options can include "allow once" (single-use permission), "allow never" (permanent denial), and "allow always" (persistent authorization).
Recent work~\cite{wu2025towards} utilizes machine learning to model user preferences of permission approvals in tool-use agents and reduces the frequency of validation prompts by predicting user decisions.

\PP{Limitations and Open Challenges}
Frequent validation prompts can overwhelm users and cause decision fatigue that undermines the intended safety benefits.
Current alert mechanisms often assume a level of security literacy that many users do not possess, causing them to either blindly approve risky actions or overreact to benign ones.
Practical systems require principled criteria for when to defer to human judgment, along with informative yet lightweight explanations that keep users engaged without overburdening them.

\subsection{Secure By Design}
\label{sec:defense:secure-by-design}
Secure-by-design defenses establish security properties at the architectural level, making agents intrinsically secure through fundamental design principles.
As secure-by-design mechanisms depend on a system's architecture and agents differ fundamentally from traditional systems, these defenses are naturally unique to agentic systems.

\subsubsection{Privilege Separation}
\label{sec:defense:privilege-separation}

Following the principle of least privilege, traditional security enforces privilege separation by assigning different privilege levels to different software components, exemplified by earlier automation attempts~\cite{brumley2004privtrans}. 
In agents, this concept extends to assigning privileges to different components and isolating these components to minimize the overall risk to the system.

\PP{Design Dimensions}
Privilege separation designs vary by \emph{separation policy} and \emph{scope}.

A \emph{separation policy} can be designed in vertical and horizontal directions.
Vertical separation divides components into hierarchical privilege levels, where higher-privilege components (e.g., trusted planners) have more authority than lower-privilege components (e.g., untrusted data processors).
Similar to kernel-user separation in operating systems, planner-processor separation designs~\cite{dual-llm,wu2024system,pfi,debenedetti2025defeating,costa2025securing,an2025ipiguard,li2025drift} isolate tool call planning (high privilege) from tool result processing (low privilege).
Memory minimization~\cite{airgap} separates the data minimizer (high privilege) from the untrusted data processing unit (low privilege).
Horizontal separation partitions the system into parallel components with equal privileges but isolated access scopes.
For example, per-application or per-functionality agents~\cite{wu2025isolategpt, li2025ace} each interact only with their dedicated tools and resources, preventing cross-domain exploitation.

The \emph{scope} of privilege separation indicates the component being separated. An LLM can be separated into planning tool calls and processing results~\cite{wu2024system,pfi,debenedetti2025defeating,costa2025securing}, preventing tool results from directly influencing planning decisions. Memory separation~\cite{airgap} creates separate memory spaces, protecting sensitive data from lower-privilege components.
Agents can separate external environments to enforce least-privilege~\cite{pfi,wu2025isolategpt,li2025ace}, leveraging sandboxing, containerization~\cite{namespace,cgroups,seccomp}, and access tokens~\cite{google-oauth,slack-oauth}.

\PP{Limitations and Open Challenges}
Current privilege separation research primarily addresses generic indirect prompt injection in simplified environments~\cite{debenedetti2024agentdojo}. It often fails to address real-world environment risks that present diverse and complex challenges requiring specialized separation strategies, such as web, file systems, and databases.
Privilege separation entails utility loss by splitting functionality across isolated components. Dividing an agent into an effective set of least-privilege components remains an open challenge. Designing efficient communication across isolated components to reduce utility loss while maintaining security guarantees presents ongoing challenges for practical deployment.

\subsubsection{Provable Security with Formal Verification}
\label{sec:defense:formal}

Traditional security methods employ formal verification to provide theoretical proofs of correctness and security~\cite{klein2009sel4,hawblitzel2014ironclad}.
Formal verification for agentic AI systems remains an emerging but crucial research frontier that aims to bridge symbolic assurance with non-symbolic behavior modeling.

\PP{Design Dimensions}
Formal verification approaches vary across \emph{formalization target} and \emph{security property}.
First, VeriSafe Agent~\cite{lee2025safeguarding} formalizes user intent into a DSL over UI state transitions, verifying that proposed GUI actions align with the user's task before execution. Formal-LLM~\cite{li2024formal} encodes developer-defined plan constraints (e.g., required tool orderings) as pushdown automata to restrict plan generation, ensuring validity and executability.
Second, various security properties are formally verified in agent systems, including predefined safety constraints~\cite{chen2025shieldagent}, alignment with user tasks~\cite{lee2025safeguarding}, and the correctness of agent behavior with respect to specified function requirements and expected outputs~\cite{li2024formal}.

\PP{Limitations and Open Challenges}
Traditional formal methods are designed for programs written in code with structured language, while LLM-based agents operate based on probabilistic models. This fundamental difference creates challenges in developing formal models that capture the stochastic behavior of agents while preserving meaningful security guarantees.
Moreover, security properties of agents that interact with various environments, such as web, file systems, or mobile applications, remain underspecified. Developing formal verification frameworks capable of addressing the full spectrum of agent security requirements largely remains an open challenge.
To develop formal verification for agentic systems, automation is critical, and frontier AI can help with the process, such as automating the specification generation~\cite{yang2024formal}.

\subsection{Identity and Access Management}
\label{sec:defense:trust}

Identity and Access Management~(IAM) encompasses identity management, access control, and  credential management to ensure authenticated entity and authorized resource access.
Traditional systems employ well-established IAM frameworks such as role-based access control~(RBAC)~\cite{sandhu1998role} and OAuth-based delegation~\cite{oauth,openid-connect}, operating with static user identities and predefined permission boundaries.
Agentic systems interact with real-world services on behalf of users, thus requiring agent-specific identities, delegation mechanisms, and dynamic access control policies that adapt to runtime contexts.

\subsubsection{Identity Management}
\label{sec:defense:identity}

Identity management ensures that each actor operates under the correct identity through user authentication and authorization.
It is a prerequisite for access control~(\autoref{sec:defense:access-control}), as proper identity authentication is required before granting access to the resource.
The identity management of agents should support delegation, auditability, and regulatory accountability, and align with existing standards~\cite{oauth,openid-connect,w3c-did}.

\PP{Design Dimensions}
Identity management varies by \emph{architecture}, \emph{scope}, and \emph{delegation model}.

First, identity management can be centralized or decentralized.
Centralized approaches~\cite{south2025authenticated,syros2025saga} rely on a central registry or identity providers~\cite{okta,composio} running on OpenID Connect~\cite{openid-connect}.
Decentralized approaches, on the other hand, use distributed verification protocols for identity verification or authentication.
For instance, agent Network Protocol~(ANP)~\cite{chang2025agent} utilizes decentralized identity authentication based on the W3C Decentralized Identifier~(DID) standard~\cite{w3c-did}, and Microsoft Verified ID~\cite{microsoft-verified-id} enables peer-to-peer identity verification.
Decentralized identifiers provide autonomy and resilience compared to centralized identity management, while requiring more complex implementations and coordination mechanisms.

Second, the scope of identity defines the principal, the entity being authenticated and authorized.
Identity can be defined at user-level, agent-level, or task-level, each providing different granularity and accountability. User-level identity ties actions directly to a human user, agent-level identity assigns distinct identities to individual agents~\cite{chan2024visibility}, and task-level identity creates short-lived identities for specific tasks or sessions to support dynamic, least-privilege operation.

Third, \emph{delegation model} governs how authority flows from users to agents. Direct delegation grants specific user permissions to agents, proxy delegation uses intermediary service or tokens to let agent act on behalf of users, and temporary delegation provides time-limited access that automatically expires to limit risk.

\PP{Limitations and Open Challenges}
Foundational questions persist about whether agent actions should be attributed to the human operator, the agent instance, or transient task identities, and the answer often varies across domains. Without standard frameworks, platforms implement incompatible credential issuance, delegation, and revocation flows that are hard to audit or federate. Robust identity management will require interoperable taxonomies and lifecycle tooling that preserve accountability while supporting seamless agent collaboration.

\subsubsection{Access Control}
\label{sec:defense:access-control}

Traditional systems protect private resources with access control by restricting access to authorized entities only.
In AI agents, an agent user's private resources reside in memory (e.g., agent usage histories and personalized knowledge bases) and the environment (e.g., file systems, cloud drives).
Recent research proposes a few methods to enforce access controls for these resources by constraining the tools and data sources the agent can access.

\PP{Design Dimensions}
Access control designs can have different \emph{mechanisms}, \emph{policies}, and \emph{resource scope}.

First, access control \emph{mechanisms} determine how permissions are enforced, such as role-based access control (RBAC)~\cite{yao2025controlnet,zhong2025honeybee}, attribute-based access control (ABAC)~\cite{amazon-bedrock}, and capability-based security systems.
Specifically, for vector database access control, access can be authorized by model activation patterns~\cite{yao2025controlnet}, outputs can be filtered~\cite{amazon-bedrock}, or the database can be partitioned to expose only accessible entries~\cite{zhong2025honeybee}.
When an agent accesses an external web service, existing mechanisms such as API key-based authentication or OAuth 2.0 protocol~\cite{oauth} can be utilized, paired with a secure delegation protocol~\cite{south2025authenticated}.
In multi-agent systems, managing access control for sub-agents is also important to prevent confused-deputy attacks~\cite{hardy1988confused}, where an agent illegally gains a privilege via another agent's capability.
For example, a recent study~\cite{syros2025saga} proposed a cryptographic protocol to enforce user-defined policies for multi-agent communications, controlling inter-agent access permissions.

Second, access control \emph{policies} define when access is granted, ranging from static permission rules to dynamic policies that adapt to changing contexts, user tasks, and environmental state.

Third, access control can target different \emph{resource scope} including agent internal memory, external databases, tool APIs, and file systems.

\PP{Limitations and Open Challenges}
Existing access control work primarily targets retrieval-augmented LLM applications with vector databases~\cite{amazon-bedrock,zhong2025honeybee,yao2025controlnet}, but agentic systems require broader coverage of diverse tools, data sources, and inter-agent interactions.
Current deployments lack adaptive policy frameworks that can dynamically adjust to evolving tasks and trust contexts, instead relying on ad hoc, non-uniform policies that create mismatches across agents and gaps when coordinating with non-agentic services~\cite{cross-agent-privilege}.
Usability challenges further exacerbate the problem, as configuration complexity leads to misconfigurations and excessive privileges even for technical users.

\subsubsection{Credential Management}
\label{sec:defense:credential}

AI agents must manage diverse credential types to interact with external services and access user resources.
These include tool API credentials (e.g., API keys and access tokens for third-party services) and environmental credentials (e.g., one-time passwords from email, session tokens).
The agent's exposure to these sensitive credentials raises significant privacy and security concerns~\cite{chatgpt-banned-italy,deepseek-banned-korea}, necessitating robust credential and secret management practices.

\PP{Design Dimensions}
Credential management approaches span \emph{confidential storage}, \emph{lifecycle management}, and \emph{credential provisioning}.

First, \emph{confidential storage} protects credentials through various storage mechanisms.
Encrypted storage protects credentials at rest, preventing unauthorized access even if the storage media is compromised.
Temporary storage minimizes exposure by maintaining credentials only for the duration of active sessions, exemplified by OpenAI's temporary chat feature~\cite{openai-temporary-chat} that prevents chat history storage and model training usage.
Dedicated credential vaults store encrypted tokens separately from agent code, so that credentials are never directly exposed to the agent~\cite{token-vault}.
Confidential computing techniques can be leveraged to securely manage credentials within hardware-based trusted execution environments~\cite{lee2020keystone,sev2020strengthening}.

Second, \emph{lifecycle management} determines how credentials are maintained over time, ranging from static credentials that persist throughout agent sessions to dynamic time-limited tokens that automatically expire, reducing exposure windows.

Third, \emph{credential provisioning} determines how agents obtain credentials, including single-sign-on (SSO) mechanisms~\cite{composio} and authenticated delegation based on OAuth 2.0~\cite{south2025authenticated} for secure credential transfer from users to agents.

\PP{Limitations and Open Challenges}
Agent stacks still rely on ad-hoc secret handling, lacking standardized practices for credential protection.
Many frameworks guide developers to store internal credentials, such as API keys, as unencrypted environment variables, increasing the risk of leakage.
Multi-agent workflows further complicate secret sharing making it difficult to maintain least privilege and traceability, and underscoring the need for coordinated credential orchestration.

\subsection{Component Hardening}
\label{sec:defense:individual}

Component hardening strengthens individual agent components, i.e., models and tools, against their specific vulnerabilities. 
It follows the principle that a system is only as secure as its weakest component.

\PP{Model Hardening}
{SecAlign}~\cite{chen2025secalign} and {StruQ}~\cite{chen2024struq} fine-tune models to consistently follow initial instructions even when faced with conflicting directives, mitigating incorrect or unintended instruction following.
Instruction-hierarchy-aware model training~\cite{wallace2024instruction,wu2024instructional} ensures that system prompts maintain priority over user input and external data.

\PP{Tool Hardening}
Extended Tool Definition Interface (ETDI)~\cite{etdi} implements cryptographically signed and versioned tool control metadata, ensuring integrity throughout the tool lifecycle to prevent tool poisoning.
MCP Context Protector~\cite{mcp-context-protector} creates an MCP proxy that enforces manual review processes and applies guardrail checks on tool descriptions and responses.
MCP Safety Audit~\cite{radosevich2025mcp} introduces systematic protocols to examine agent tools, identifying potentially exploitable behaviors from malicious logic or misleading descriptions.
MCIP~\cite{jing2025mcip} enhances MCP with observability and an LLM fine-tuned to detect threats in MCP usage.
Anthropic's Connectors Directory~\cite{anthropic-connectors} maintains a curated repository of trusted tools with reviewed descriptions, functionality, and safety policies.


\PP{Limitations and Open Challenges}
Current component hardening approaches focus on simplified threat scenarios that do not reflect the complexity of real-world agent systems. Model hardening techniques like instruction hierarchy fine-tuning primarily address simple scenarios involving system prompts, user prompts, and tool results, failing to address sophisticated attacks such as shadowing attacks, where malicious tool results override other tool results. The role of each component and appropriate threat models for comprehensive hardening remain unclear.

\subsection{Defense Design Principles}
\label{sec:defense:principles}

Effective agent security requires multiple complementary defense mechanisms working together rather than relying on any single approach.
Three fundamental security principles from traditional security guide secure agent defense design.
As discussed in \autoref{sec:agent}, AI agents are hybrid software systems that combine LLMs with traditional software components, inheriting the same security concerns that motivated classical defense principles. Agents further amplify the need for these principles due to their autonomous decision-making, heterogeneous trust boundaries, and complex multi-step execution.
We note that additional principles can also be applied to agents (e.g., fail-safe defaults and economy of mechanisms).

\PP{Defense-in-Depth}
Defense mechanisms complement each other by operating at different stages and targeting different attack vectors. Input guardrails~(\autoref{sec:defense:input}) provide first-line protection by filtering malicious inputs, while output guardrails~(\autoref{sec:defense:output}) serve as last-line defense by sanitizing agent outputs. Information flow control~(\autoref{sec:defense:flow}) and monitoring~(\autoref{sec:defense:monitor}) provide continuous runtime protection throughout agent execution, while access control~(\autoref{sec:defense:access-control}) ensures proper authentication and authorization.
Secure-by-design approaches like privilege separation~(\autoref{sec:defense:privilege-separation}) establish fundamental architectural protections. Component hardening~(\autoref{sec:defense:individual}) strengthens individual elements, and human-in-the-loop validation~(\autoref{sec:defense:human}) provides user oversight for critical decisions.
However, layering defenses can also introduce \emph{emergent misalignment}~\cite{betley2025emergent}, where one mechanism inadvertently weakens another. For example, a sanitizer may strip safety instructions relied upon by a downstream guardrail. Defense-in-depth therefore requires coordinated design across the full agent stack.

\PP{Principle of Least Privilege}
Agents should operate with the minimum necessary permissions and access rights. This principle is implemented through privilege separation techniques~(\autoref{sec:defense:privilege-separation}) that isolate agent components and restrict tool access to only what is required for specific tasks, as well as identity management~(\autoref{sec:defense:identity}) that defines appropriate access scopes for agents.

\PP{Complete Mediation}
All access to sensitive resources should be verified and authorized. This principle is reflected in comprehensive monitoring systems~(\autoref{sec:defense:monitor}), access control mechanisms~(\autoref{sec:defense:access-control}), and identity management~(\autoref{sec:defense:identity}) that verify every agent interaction with protected resources.

\DeclareRobustCommand{\supportfull}{%
  \tikz[baseline=-0.6ex]{\fill (0,0) circle (1ex);}
}
\DeclareRobustCommand{\supportpartial}{%
  \tikz[baseline=-0.6ex]{%
    \fill (0,0) circle (1ex);
    \path[fill=white] (-1.2ex,-1.2ex) rectangle (0,1ex);
    \draw (0,0) circle (1ex);
  }%
}

\begin{table*}[t]
  \footnotesize
  \centering
  \caption{Defense coverage in real-world agents. Defense techniques include input guardrail~(Input), output guardrail~(Output), access control~(Access), information flow control and taint tracking~(IFC), monitoring~(Monitor), human-in-the-loop~(HITL), privilege separation~(Priv), formal verification~(Formal), identity management~(ID), and credential management~(Cred). \supportfull{} indicates full support and \supportpartial{} indicates partial support. }
  \begin{adjustbox}{max width=\linewidth}
  \begin{tabular}{llcccccccccc}
    \toprule
    \textbf{Category} & \textbf{Agent} & \textbf{Input} & \textbf{Output} & \textbf{Access} & \textbf{IFC} & \textbf{Monitor} & \textbf{HITL} & \textbf{Priv} & \textbf{Formal} & \textbf{ID} & \textbf{Cred} \\
    \midrule
    Coding & Codex~v0.53.0~\cite{codex} &  & \supportpartial & \supportfull &  & \supportpartial & \supportfull &  &  &  & \\
    Coding & Gemini CLI~v0.13.0~\cite{gemini-cli} &  & \supportpartial &  &  & \supportpartial & \supportfull &  &  &  & \\
    Coding & OpenHands~v0.59.0~\cite{openhands} &  & \supportfull &  &  & \supportpartial & \supportfull & \supportfull &  &  &  \\
    \midrule
    Web & Browser Use~v0.9.0~\cite{browser-use} & \supportpartial & \supportpartial &  &  & \supportpartial &  &  &  &  & \supportpartial \\
    Web & Nanobrowser~v0.1.12~\cite{nanobrowser} & \supportfull & \supportpartial &  &  & \supportpartial &  & \supportfull &  &  &  \supportpartial\\
    Web & Skyvern~v0.2.20~\cite{skyvern} & \supportpartial & \supportpartial &  &  & \supportpartial &  &  &  &  & \supportpartial   \\
    \bottomrule
\end{tabular}
  \end{adjustbox}
\label{tab:case-study}
\end{table*}
\section{Securing Real-World Agents}
\label{sec:case-study}


We analyze six open-source agents to illustrate how real-world agentic systems combine defenses across the dimensions in \autoref{sec:defense}.
We focus on system-level defenses rather than component hardening.
\autoref{tab:case-study} highlights which defense classes each agent enables.
We consider a defense partially supported when the system provides incomplete coverage (e.g., only protects against a subset of threats) or requires non-negligible manual effort (e.g., manual configuration or curation).
A defense is fully supported when the protection is automated and provides comprehensive coverage, even if its accuracy is imperfect.
Note that this case study represents the agents' current status; these agents are actively evolving and they continue to update and strengthen their defenses.

\subsection{Coding agents}

\PP{General Coding Agent Defenses}
Coding agents typically operate within directories that users trust.
Nevertheless, threats remain from multiple sources: the LLM may hallucinate and generate incorrect code, the model itself could be compromised by backdoors~(\vectormodel), or seemingly trusted inputs (e.g., user input, code repositories, documentation) may contain hidden malicious instructions~(\vectorindirect,\vectordirect) without the user's awareness.

To defend against such threats, coding agents prioritize constraining agent actions over sanitizing inputs.
For instance, they gate AI-generated filesystem operations and shell commands through output guardrails~(\autoref{sec:defense:output}) and access control~(\autoref{sec:defense:access-control}) instead of filtering prompts.
They also lean heavily on human-in-the-loop validation~(\autoref{sec:defense:human}) for sensitive operations because, while these actions can disrupt a user's machine, their safety depends on context. 
Implementations vary in how they define sensitive actions and enforce access control. 
Current monitoring~(\autoref{sec:defense:monitor}) support is partial in all coding agents. They collect logs of agent actions (e.g., tool calls, file modifications) using services like OpenTelemetry~\cite{opentelemetry} or PostHog~\cite{posthog}, enabling post-hoc manual review, but lack detections for suspicious patterns or anomalies.

\PP{Codex}
Codex~\cite{codex} combines access control with human-in-the-loop validation to secure AI-suggested file patches and shell commands, while also providing partial output guardrails and monitoring.
It implements access control through path restrictions and privilege escalation controls. For file patches, the system requests user approval when the target file is outside writable paths (e.g., not in the working directory). For shell commands, the agent executes them in a sandbox by default, which restricts access to the working directory with no network access~\cite{codex-sandbox}. The model can request elevated privileges to run a command outside the sandbox when necessary. Such escalation requires user approval, providing human-in-the-loop control over potentially dangerous operations. This design ensures automatic agent actions run under containment while allowing controlled privilege escalation for legitimate use cases. 

\PP{Gemini CLI}
Gemini CLI~\cite{gemini-cli} relies primarily on human-in-the-loop validation for command execution, complemented by partial output guardrails and monitoring.
It applies output guardrails to file access and shell commands, restricting file reads to workspace directories. For shell commands, it maintains allow and deny lists, prompting the user before running anything unlisted. Users can cache decisions, and the agent parses compound commands into individual components so each decision is stored independently, reducing redundant user prompts. As a rule-based guardrail, coverage remains partial. The underlying environment ultimately depends on user-controlled permissions, with no additional OS-level access control. Gemini CLI encourages running the agent inside containers (e.g., Docker) to strengthen access control by isolating the entire agent~\cite{gemini-sandbox}, though setting up and maintaining that sandbox adds nontrivial overhead.

\PP{OpenHands}
OpenHands~\cite{openhands} takes a multi-layered approach, combining output guardrails with human-in-the-loop validation and privilege separation, along with partial monitoring support.
It implements a more active output guardrail by asking the LLM to emit a \texttt{security\_risk} score with every tool decision to detect high-risk tool usage in a context-sensitive manner. Those high-risk tool calls cannot proceed without user consent, preserving human-in-the-loop control. Like Gemini CLI, it skips per-command sandboxing and recommends containerization to strengthen access control by isolating the entire agent. For privilege separation, OpenHands employs a multi-agent architecture where each agent is granted access to different tools and capabilities (e.g., a coding agent has file system access while a web browsing agent has network access). This separation limits the impact of compromising any individual agent and supports secure delegation schemes across agent boundaries.

\PP{Future Directions}
Effective security for coding agents requires defense-in-depth with multiple complementary mechanisms working together. Current implementations have significant gaps in both coverage and effectiveness.

First, coding agents lack several critical defenses. They should deploy input guardrails~(\autoref{sec:defense:input}) that validate workspaces and user prompts to surface poisoned data before agents act on it, and filter web-retrieved content for prompt-injection payloads before it reaches the model. Information flow control~(\autoref{sec:defense:flow}) could track how untrusted data influences tool calls, preventing malicious instructions from compromising agent decisions. Identity and credential management~(\autoref{sec:defense:identity}, \autoref{sec:defense:credential}) would enable proper authentication and secure storage of API keys and access tokens.

Second, existing partial defenses need significant strengthening. Access control~(\autoref{sec:defense:access-control}) should move beyond all-or-nothing approvals to seamless, fine-grained permissions that grant only the minimum necessary access across diverse development tools. Human-in-the-loop validation~(\autoref{sec:defense:human}) needs richer contextual information, such as expected side effects or comparisons with past approvals, to help users avoid decision fatigue and make informed authorizations. Monitoring~(\autoref{sec:defense:monitor}) should pair existing logs with automated detectors that identify domain-specific risks and privilege-escalation attempts so users can intervene before damage occurs.


\subsection{Web agents}
Web agents autonomously perform web tasks that range from summarizing pages to navigating URLs, clicking buttons, and completing forms. As web agents receive arbitrary content from diverse sources, they inherently process untrusted data even when tasks require sensitive inputs such as personal identifiers, payment details, or authenticated workflows (e.g., accessing cloud files, sending email, or making purchases).

\PP{General Web Agent Defenses}
Web agents are particularly vulnerable to indirect prompt injection attacks~(\vectorindirect), where malicious instructions embedded in web pages hijack agent behavior.
Unlike coding agents that primarily operate in trusted workspaces, web agents continuously process untrusted external inputs from vast and diverse web sources.
Moreover, web agents often access the web with user authorization, handling sensitive private data and credentials (e.g., accessing cloud files, reading emails, making purchases), which makes them high-value targets.
Additional threats include compromised models~(\vectormodel), hallucinations, and direct attacks from user prompts~(\vectordirect).

Given the vast untrusted input surface and access to sensitive data, web agents commonly employ four defense mechanisms: input guardrails~(\autoref{sec:defense:input}), output guardrails~(\autoref{sec:defense:output}), credential management~(\autoref{sec:defense:credential}), and monitoring~(\autoref{sec:defense:monitor}).
Input guardrails filter malicious content from web pages by using techniques such as domain allow and deny lists to control which sites agents can visit, and filtering page elements or blocklisted domains to prevent malicious instructions from reaching the model.
Output guardrails constrain agent actions by preventing navigation to sensitive URLs such as local-network hosts, raw IP addresses, and browser configuration pages like \texttt{chrome://settings}, limiting server-side request forgery attempts.
Credential management protects sensitive user data through various techniques such as redacting or replacing secrets before they reach LLM providers or untrusted domains.
For monitoring, web agents emit browsing telemetry for post-hoc review, yet none of the surveyed systems pair these logs with automated detection.
However, most defenses provide only partial protection, with manually curated controls and incomplete coverage.

\PP{Browser-use}
Browser-use~\cite{browser-use} provides support for input guardrails, output guardrails, credential management, and monitoring.
For input guardrails, it applies ad-block rules to strip advertising and other unwanted elements from incoming pages, which offers limited coverage against malicious prompts. For credential management, it protects secrets by replacing them with placeholders before data reaches third-party LLM providers or untrusted domains, using user-defined mappings that specify which secrets may be revealed to which sites.

\PP{Nanobrowser}
Nanobrowser~\cite{nanobrowser} implements input guardrails and privilege separation, along with partial support for output guardrails, credential management, and monitoring.
It strengthens input guardrails by inserting delimiters and guard prompts that keep user instructions distinct from retrieved page content, mitigating indirect prompt injection attacks. For credential management, it redacts sensitive data, such as Social Security numbers, credit card details, and email addresses, protecting those values from LLM providers and untrusted sites. Such data are detected using regular expression rules. For privilege separation, Nanobrowser splits responsibilities between planner and navigator agents with different permissions, so that compromising the navigator cannot directly corrupt the overall agent plan.

\PP{Skyvern}
Skyvern~\cite{skyvern} provides support for input guardrails, output guardrails, credential management, and monitoring, with a particular focus on credential protection.
For credential management, it specializes in protecting one-time passwords~(OTPs) for automated authorization tasks. It detects OTPs at runtime with regular expressions, swaps them with placeholders to keep the secrets from LLM providers and untrusted websites, and restores the original values only to the authentication form. While this raises the bar for OTP exfiltration attacks, the regex-based approach provides incomplete coverage and does not extend to other credential types.

\PP{Future Directions}
Like coding agents, securing web agents requires defense-in-depth with multiple complementary mechanisms. Today's web agents remain early-stage prototypes with utility features still maturing and defenses that are simple and fragile.

First, web agents lack several critical defenses entirely. Information flow control~(\autoref{sec:defense:flow}) could track how untrusted web content influences agent decisions and prevent data exfiltration to malicious domains. Identity management~(\autoref{sec:defense:identity}) would enable proper authentication when agents act on behalf of users across multiple web services. Human-in-the-loop validation~(\autoref{sec:defense:human}) could provide user oversight for high-impact web actions such as purchases or data sharing.

Second, existing partial defenses need significant strengthening. Input guardrails~(\autoref{sec:defense:input}) should move beyond manual domain allow and deny lists to automated domain reputation models and contextual filtering, reducing user burden and addressing risks such as expired domains~\cite{roth2020complex}. Structural protections combining privilege separation~(\autoref{sec:defense:privilege-separation}), taint tracking, and model-level input guardrails can neutralize malicious instructions before they reach planners. Monitoring~(\autoref{sec:defense:monitor}) should pair existing telemetry with real-time detectors that flag or halt suspicious actions and escalate to human validation when necessary. Credential management~(\autoref{sec:defense:credential}) requires adaptive and privacy-preserving detection techniques to better protect diverse credential types beyond simple regex-based approaches.

\section{Detailed Case Study: AutoGPT}
\label{sec:case:autogpt}

AutoGPT~\cite{autogpt} is one of the most widely used open-source autonomous agents with over 180k GitHub stars.
It exposes a broad set of tools enabling LLM interaction with heterogeneous environments, including the Internet, local files, and execution interfaces (e.g., command line).
In this section, we analyze multiple versions of AutoGPT since v0.4.3, track their real-world vulnerability reports, and evaluate implemented defenses.


\subsection{Tools and Execution Environments}
\label{subsec:case:autogpt:env}

AutoGPT equips agents with retrieval tools for information gathering, execution tools for system-level operations.
These tools enable AutoGPT to interact with diverse external environments.

\PP{Retrieval Tools.}
Agents can search the web with \texttt{google\_search}, fetch webpage content through \texttt{browse\_website}, navigate local files using \texttt{read\_file} and \texttt{search\_files}, and access historical context via \texttt{load\_from\_memory}.

\PP{Execution Tools.}
AutoGPT grants direct system access through \texttt{execute\_shell} for arbitrary bash commands, \texttt{file\_manipulation} for modifying workspace contents, and \texttt{execute\_python\_code} for dynamic code execution.

\PP{External Environments.}
AutoGPT interacts with the web (Internet access via \texttt{google\_search}), the computer (filesystem read/write access), and domain-specific environments (Python interpreter, shell, OS-level interfaces).

\subsection{Real-world Vulnerabilities in AutoGPT}
\label{subsec:case:autogpt:attack}

We study five representative CVE vulnerabilities since 2023 and map them to our risk taxonomy.

\PP{A. Docker-Compose Overwrite (CVE-2023-37273)}
The \texttt{docker-compose.yml} file of the project lacks write protection, allowing malicious LLM outputs to overwrite container configurations.
Attackers embed malicious instructions in external content (e.g., a web page fetched by the agent), which hijack the LLM into calling \texttt{execute\_python\_code} to overwrite the configuration file~(\riskwronginst).
When AutoGPT restarts, it executes the malicious container, leading to container escape and host compromise~(\riskintegrity).

\PP{B. Path Traversal (CVE-2023-37274)}
Unsanitized basename parameters allow path traversal attacks that write files outside the sandbox.
Attackers inject instructions into external content that trick the LLM into calling \texttt{execute\_python\_code} with a traversal path such as \texttt{../../main.py}, overwriting critical AutoGPT source files~(\riskwronginst, \riskintegrity).
When AutoGPT restarts, these modified files execute, achieving persistent arbitrary code execution.
The overwritten files may also expose sensitive source code or configuration data~(\riskconfidentiality).

\PP{C. ANSI Escape Sequence Deception (CVE-2023-37275)}
ANSI escape sequences are special control codes interpreted by terminals to perform actions such as moving the cursor, clearing the screen, or changing text color.
When AutoGPT fetches external web content via \texttt{browse\_website}, it passes the retrieved content—including any embedded ANSI codes—directly to the console without sanitization.
An attacker crafts a malicious web page embedding JSON-encoded ANSI escape sequences.
The sequences are not instructions that hijack the LLM; rather, they flow as unsanitized data through the agent pipeline and are rendered by the terminal~(\riskunsafedataflow).
The spoofed console output can conceal executed commands or trick the human operator into approving malicious actions, silently hijacking the agent's behavior~(\riskintegrity).

\PP{D. Cross-Site Request Forgery (CVE-2024-1879)}
Missing CSRF protection and permissive CORS settings allow authenticated API requests from malicious webpages.
An attacker crafts a webpage that, when visited by an authenticated user, silently triggers agent actions through cross-origin requests.
This enables unauthorized command execution and data exfiltration~(\riskconfidentiality, \riskintegrity).

\PP{E. OS Command Injection (CVE-2024-1881)}
AutoGPT validates shell commands using an allowlist that checks only the first token.
This approach blocks individual dangerous commands but fails to detect operator-chained payloads or multiple commands in a single line.
Attackers inject instructions into external content that manipulate the LLM to generate commands with chaining operators (e.g., \texttt{ls \&\& rm -rf /}, \texttt{cat file; curl attacker.com})~(\riskwronginst).
Even without an attacker, the LLM may spontaneously generate chained shell commands for complex tasks, bypassing the first-token check~(\riskhallucination).
The executor runs these multi-command payloads verbatim, enabling arbitrary command execution, data exfiltration, and filesystem compromise~(\riskconfidentiality, \riskintegrity).

\subsection{Defenses in AutoGPT}
\label{subsec:case:autogpt:defense}

AutoGPT has deployed patches across multiple versions to address the known vulnerabilities discussed above.
\autoref{tab:autogpt-defense} summarizes the defense landscape per CVE: the risks each vulnerability exploits, the defense mechanism applied, which risks the patch mitigates, which remain open, and what defenses are still missing.
Notably, all patches target downstream consequences (access control and output sanitization) rather than the upstream causes, leaving indirect prompt injection~(\riskwronginst) and unsafe data flow~(\riskunsafedataflow) unaddressed at their source.

\begin{table*}[t]
  \footnotesize
  \centering
  \caption{Defense analysis of AutoGPT CVEs. For each vulnerability, we list the exploited risks, the defense mechanism deployed in the patch, which risks are mitigated, which remain open, and which defense categories are missing.}
  \begin{adjustbox}{max width=\linewidth}
  \begin{tabular}{llllll}
    \toprule
    \textbf{CVE} & \textbf{Exploited Risks} & \textbf{Defense Mechanism} & \textbf{Risks Mitigated} & \textbf{Risks Still Open} & \textbf{Missing Defenses} \\
    \midrule
    A.~Docker-Compose & \riskwronginst, \riskintegrity & Access control & \riskintegrity & \riskwronginst & Input guardrails, \\
    \quad CVE-2023-37273 & & (read-only mounts) & & & Information flow control \\
    \midrule
    B.~Path Traversal & \riskwronginst, \riskconfidentiality, \riskintegrity & Output guardrail & \riskintegrity & \riskwronginst, \riskconfidentiality & Input guardrails, \\
    \quad CVE-2023-37274 & & (path canonicalization) & & & Information flow control \\
    \midrule
    C.~ANSI Escape & \riskunsafedataflow, \riskintegrity & Output guardrail & \riskunsafedataflow~(partial) & \riskunsafedataflow~(incomplete) & Information flow control \\
    \quad CVE-2023-37275 & & (regex ANSI filter) & & & \\
    \midrule
    D.~CSRF & \riskconfidentiality, \riskintegrity & Access control & \riskconfidentiality, \riskintegrity~(partial) & \riskconfidentiality, \riskintegrity & Identity management, \\
    \quad CVE-2024-1879 & & (CSRF tokens, CORS) & & (localhost bypass) & Monitoring \\
    \midrule
    E.~Command Injection & \riskwronginst, \riskhallucination, & Output guardrail & \riskintegrity~(partial) & \riskwronginst, \riskhallucination, & Output guardrails, \\
    \quad CVE-2024-1881 & \riskconfidentiality, \riskintegrity & (first-token allowlist) & & \riskconfidentiality & Human-in-the-loop, \\
    & & & & & Privilege separation, \\
    & & & & & Monitoring, Formal verification \\
    \bottomrule
  \end{tabular}
  \end{adjustbox}
  \label{tab:autogpt-defense}
\end{table*}

\PP{A. Docker-Compose Overwrite (CVE-2023-37273)}
Versions after 0.4.3 use read-only mounts and restrict permissions on configuration files.
This mitigates the integrity impact~(\riskintegrity) by preventing overwrites of \texttt{docker-compose.yml}, but does not address the indirect prompt injection~(\riskwronginst) that triggers the overwrite attempt.
Input guardrails~(\autoref{sec:defense:input}) should scan fetched web content for prompt injection, and information flow control~(\autoref{sec:defense:flow}) should prevent tainted LLM outputs from reaching \texttt{execute\_python\_code}.

\PP{B. Path Traversal (CVE-2023-37274)}
Versions after 0.4.3 canonicalize paths and filter traversal patterns like \texttt{../} in the \texttt{agent.workspace.get\_path()} function.
This blocks basic traversal attacks~(\riskintegrity) when the workspace root is properly configured, but the indirect prompt injection vector~(\riskwronginst) that causes the LLM to generate traversal paths remains unmitigated.
Input guardrails and information flow control~(\autoref{sec:defense:flow}) should block the injection at its source and track taint from web content to filesystem operations.

\PP{C. ANSI Escape Sequence Deception (CVE-2023-37275)}
Versions after 0.4.3 apply rule-based sanitization to filter escape sequences from model outputs.
This partially addresses the unsafe data flow~(\riskunsafedataflow), but cannot cover all escape sequence variants and may be bypassed through novel encoding schemes or lesser-known control codes.
Information flow control~(\autoref{sec:defense:flow}) should treat all data originating from external web sources as untrusted and enforce sanitization at every output boundary, including the terminal, not just the LLM context.

\PP{D. Cross-Site Request Forgery (CVE-2024-1879)}
Versions after 0.5.1 add CSRF tokens and enforce strict CORS policies that trust only localhost ports.
This prevents most cross-origin attacks but leaves open attacks from malicious browser extensions or local applications that can access localhost endpoints.
Identity management~(\autoref{sec:defense:trust}) should implement OAuth-based authorization instead of relying solely on token validation.
Monitoring~(\autoref{sec:defense:monitor}) should track cross-origin patterns to catch attacks before data theft.

\PP{E. OS Command Injection (CVE-2024-1881)}
Current versions maintain a command allowlist that validates the first token of each input.
This blocks individual dangerous commands but does not prevent operator-based chaining or multi-command payloads.
The defense remains incomplete: neither the indirect prompt injection~(\riskwronginst) that manipulates the LLM nor the hallucination risk~(\riskhallucination) that produces chained commands spontaneously is addressed.
Output guardrails~(\autoref{sec:defense:output}) should parse shell syntax to catch chained payloads regardless of whether they originate from an attacker or from the LLM's own generation.
Human-in-the-loop validation~(\autoref{sec:defense:human}) should display command risks so users can judge high-risk operations.
Privilege separation~(\autoref{sec:defense:privilege-separation}) should isolate shell execution with minimal privileges per risk level.
Monitoring~(\autoref{sec:defense:monitor}) should log command provenance, and formal verification~(\autoref{sec:defense:formal}) should restrict operations to pre-approved command templates.
\section{Conclusion}
\label{sec:conclusion}

This paper presents an overview of the attack and defense landscape for AI agents, together with an in-depth analysis of AI agent risks, security goals, defense dimensions, case studies, and open challenges.
Our survey reveals that while agentic AI security research has made significant progress in mapping the problem space, practical and general-purpose defenses remain largely elusive.
Critical directions for the field include realistic evaluation frameworks that bridge research and production, composable defenses that avoid emergent misalignment, standardized agent identity and access control, and adaptive defenses that balance security with usability.
Our SoK can serve as a guide for building secure agents and point out meaningful directions for future research.



\bibliographystyle{ACM-Reference-Format}
\bibliography{ref}

@inproceedings{contextual-integrity,
  title={Privacy and contextual integrity: Framework and applications},
  author={Barth, Adam and Datta, Anupam and Mitchell, John C and Nissenbaum, Helen},
  booktitle={2006 IEEE symposium on security and privacy (S\&P'06)},
  pages={15--pp},
  year={2006},
  organization={IEEE}
}

@misc{browser-use,
  title = {Browser use},
  author = {Browser use},
  year = {2024},
  note= {\url{https://browser-use.com/}},
}

@misc{nanobrowser,
  title = {Nanobrowser},
  author = {Nanobrowser},
  year = {2025},
  note = {\url{https://nanobrowser.ai/}},
}

@misc{skyvern,
  title = {Skyvern},
  author = {{Skyvern}},
  year = {2025},
  note = {\url{https://www.skyvern.com/}},
}

@misc{codex,
  title = {OpenAI Codex},
  author = {{OpenAI}},
  year = {2025},
  note = {\url{https://openai.com/codex/}},
}

@misc{openhands,
  title = {OpenHands},
  author = {All Hands AI},
  year = {2025},
  note = {\url{https://openhands.dev/}},
}

@misc{gemini-cli,
  title = {Gemini CLI},
  author = {Google},
  year = {2025},
  note = {\url{https://docs.cloud.google.com/gemini/docs/codeassist/gemini-cli}},
}

@article{zou2023universal,
  title={Universal and transferable adversarial attacks on aligned language models},
  author={Zou, Andy and Wang, Zifan and Carlini, Nicholas and Nasr, Milad and Kolter, J Zico and Fredrikson, Matt},
  journal={arXiv preprint arXiv:2307.15043},
  year={2023}
}

@article{universal-prompt,
  title={Automatic and universal prompt injection attacks against large language models},
  author={Liu, Xiaogeng and Yu, Zhiyuan and Zhang, Yizhe and Zhang, Ning and Xiao, Chaowei},
  journal={arXiv preprint arXiv:2403.04957},
  year={2024}
}

@article{agentpoison,
  title={Agentpoison: Red-teaming llm agents via poisoning memory or knowledge bases},
  author={Chen, Zhaorun and Xiang, Zhen and Xiao, Chaowei and Song, Dawn and Li, Bo},
  journal={Advances in Neural Information Processing Systems},
  volume={37},
  pages={130185--130213},
  year={2024}
}

@inproceedings{agentvigil,
  title={AgentVigil: Generic Black-Box Red-teaming for Indirect Prompt Injection against LLM Agents},
  author={Wang, Zhun and Siu, Vincent and Ye, Zhe and Shi, Tianneng and Nie, Yuzhou and Zhao, Xuandong and Wang, Chenguang and Guo, Wenbo and Song, Dawn},
  booktitle={Findings of the Association for Computational Linguistics: EMNLP},
  pages={23159--23172},
  year={2025},
  organization={Association for Computational Linguistics}
}

@article{llama-firewall,
  title={Llamafirewall: An open source guardrail system for building secure ai agents},
  author={Chennabasappa, Sahana and Nikolaidis, Cyrus and Song, Daniel and Molnar, David and Ding, Stephanie and Wan, Shengye and Whitman, Spencer and Deason, Lauren and Doucette, Nicholas and Montilla, Abraham and others},
  journal={arXiv preprint arXiv:2505.03574},
  year={2025}
}

@inproceedings{nemo-guardrails,
  title={Nemo guardrails: A toolkit for controllable and safe llm applications with programmable rails},
  author={Rebedea, Traian and Dinu, Razvan and Sreedhar, Makesh Narsimhan and Parisien, Christopher and Cohen, Jonathan},
  booktitle={Proceedings of the 2023 conference on empirical methods in natural language processing: system demonstrations},
  pages={431--445},
  year={2023}
}

@misc{safe-browsing,
  title={Google Safe Browsing},
  author={Google},
  year={2025},
  note={\url{https://safebrowsing.google.com/}}
}

@article{autogpt,
  title={Auto-gpt for online decision making: Benchmarks and additional opinions},
  author={Yang, Hui and Yue, Sifu and He, Yunzhong},
  journal={arXiv preprint arXiv:2306.02224},
  year={2023}
}

@misc{mcp,
  title={Introducing the Model Context Protocol},
  author={Anthropic},
  year={2024},
  note={\url{https://www.anthropic.com/news/model-context-protocol}}
}

@misc{mcp-server,
  title={Model Context Protocol servers},
  author={modelcontextprotocol},
  year={2025},
  note={\url{https://github.com/modelcontextprotocol/servers}}
}

@article{gorilla,
  title={Gorilla: Large language model connected with massive apis},
  author={Patil, Shishir G and Zhang, Tianjun and Wang, Xin and Gonzalez, Joseph E},
  journal={Advances in Neural Information Processing Systems},
  volume={37},
  pages={126544--126565},
  year={2024}
}

@inproceedings{lee2023prompted,
  title={Prompted LLMs as Chatbot Modules for Long Open-domain Conversation},
  author={Lee, Gibbeum and Hartmann, Volker and Park, Jongho and Papailiopoulos, Dimitris and Lee, Kangwook},
  booktitle={Findings of the Association for Computational Linguistics: ACL 2023},
  pages={4536--4554},
  year={2023}
}

@article{lewis2020retrieval,
  title={Retrieval-augmented generation for knowledge-intensive nlp tasks},
  author={Lewis, Patrick and Perez, Ethan and Piktus, Aleksandra and Petroni, Fabio and Karpukhin, Vladimir and Goyal, Naman and K{\"u}ttler, Heinrich and Lewis, Mike and Yih, Wen-tau and Rockt{\"a}schel, Tim and others},
  journal={Advances in neural information processing systems},
  volume={33},
  pages={9459--9474},
  year={2020}
}

@article{toolformer,
  title={Toolformer: Language models can teach themselves to use tools},
  author={Schick, Timo and Dwivedi-Yu, Jane and Dessi, Roberto and Raileanu, Roberta and Lomeli, Maria and Hambro, Eric and Zettlemoyer, Luke and Cancedda, Nicola and Scialom, Thomas},
  journal={Advances in Neural Information Processing Systems},
  volume={36},
  pages={68539--68551},
  year={2023}
}

@misc{workflow-agent,
  title={Workflow Agent},
  author={Agent Development Kit},
  year={2025},
  note={\url{https://google.github.io/adk-docs/agents/workflow-agents/}}
}

@misc{human-in-the-loop,
  title={Human-in-the-loop},
  author={langgraph},
  year={2025},
  note={url{https://langchain-ai.github.io/langgraph/concepts/human-in-the-loop/}}
}

@misc{openai-memory,
  title={Memory and new controls for ChatGPT},
  author={OpenAI},
  year={2025},
  note={\url{https://openai.com/index/memory-and-new-controls-for-chatgpt/}}
}

@article{imprompter-attack,
  title={Imprompter: Tricking LLM Agents into Improper Tool Use},
  author={Fu, Xiaohan and Li, Shuheng and Wang, Zihan and Liu, Yihao and Gupta, Rajesh K and Berg-Kirkpatrick, Taylor and Fernandes, Earlence},
  journal={arXiv preprint arXiv:2410.14923},
  year={2024}
}

@article{zhong2025rtbas,
  title={Rtbas: Defending llm agents against prompt injection and privacy leakage},
  author={Zhong, Peter Yong and Chen, Siyuan and Wang, Ruiqi and McCall, McKenna and Titzer, Ben L and Miller, Heather and Gibbons, Phillip B},
  journal={arXiv preprint arXiv:2502.08966},
  year={2025}
}

@misc{pi_against_gpt3,
  title = {{Prompt injection attacks against GPT-3}},
  howpublished = "\url{https://simonwillison.net/2022/Sep/12/prompt-injection/}",
  author = {Simon Willison},
  year={2022}
}

@misc{delimiters_url,
  title = {{Delimiters won’t save you from prompt injection}},
  howpublished = "\url{https://simonwillison.net/2023/May/11/delimiters-wont-save-you}",
  author = {Simon Willison},
  year={2023}
}

@inproceedings{perez2022ignore,
    author = {Perez, Fábio and Ribeiro, Ian},
    title = {Ignore Previous Prompt: Attack Techniques For Language Models},
    booktitle = {NeurIPS ML Safety Workshop},
    year = {2022}
}

@inproceedings{liu2024formalizing,
  title={Formalizing and benchmarking prompt injection attacks and defenses},
  author={Liu, Yupei and Jia, Yuqi and Geng, Runpeng and Jia, Jinyuan and Gong, Neil Zhenqiang},
  booktitle={33rd USENIX Security Symposium (USENIX Security 24)},
  pages={1831--1847},
  year={2024}
}

@inproceedings{debenedetti2024agentdojo,
  title={AgentDojo: A Dynamic Environment to Evaluate Prompt Injection Attacks and Defenses for LLM Agents},
  author={Debenedetti, Edoardo and Zhang, Jie and Balunovic, Mislav and Beurer-Kellner, Luca and Fischer, Marc and Tram{\`e}r, Florian},
  booktitle={The Thirty-eight Conference on Neural Information Processing Systems Datasets and Benchmarks Track},
  year={2024}
}

@inproceedings{greshake2023not,
  title={Not what you've signed up for: Compromising real-world llm-integrated applications with indirect prompt injection},
  author={Greshake, Kai and Abdelnabi, Sahar and Mishra, Shailesh and Endres, Christoph and Holz, Thorsten and Fritz, Mario},
  booktitle={Proceedings of the 16th ACM Workshop on Artificial Intelligence and Security},
  pages={79--90},
  year={2023}
}

@article{hou2025model,
  title={Model context protocol (mcp): Landscape, security threats, and future research directions},
  author={Hou, Xinyi and Zhao, Yanjie and Wang, Shenao and Wang, Haoyu},
  journal={arXiv preprint arXiv:2503.23278},
  year={2025}
}

@article{fu2023misusing,
  title={Misusing tools in large language models with visual adversarial examples},
  author={Fu, Xiaohan and Wang, Zihan and Li, Shuheng and Gupta, Rajesh K and Mireshghallah, Niloofar and Berg-Kirkpatrick, Taylor and Fernandes, Earlence},
  journal={arXiv preprint arXiv:2310.03185},
  year={2023}
}

@article{wu2024agentattack,
  title={Adversarial Attacks on Multimodal Agents},
  author={Wu, Chen Henry and Koh, Jing Yu and Salakhutdinov, Ruslan and Fried, Daniel and Raghunathan, Aditi},
  journal={arXiv preprint arXiv:2406.12814},
  year={2024}
}

@inproceedings{liao2024eia,
  title={EIA: ENVIRONMENTAL INJECTION ATTACK ON GENERALIST WEB AGENTS FOR PRIVACY LEAKAGE},
  author={Liao, Zeyi and Mo, Lingbo and Xu, Chejian and Kang, Mintong and Zhang, Jiawei and Xiao, Chaowei and Tian, Yuan and Li, Bo and Sun, Huan},
  booktitle={The Thirteenth International Conference on Learning Representations},
  year={2025}
}

@inproceedings{schulhoff2023ignore,
    title = "Ignore This Title and {H}ack{AP}rompt: Exposing Systemic Vulnerabilities of {LLM}s Through a Global Prompt Hacking Competition",
    author = "Schulhoff, Sander  and
      Pinto, Jeremy  and
      Khan, Anaum  and
      Bouchard, Louis-Fran{\c{c}}ois  and
      Si, Chenglei  and
      Anati, Svetlina  and
      Tagliabue, Valen  and
      Kost, Anson  and
      Carnahan, Christopher  and
      Boyd-Graber, Jordan",
    editor = "Bouamor, Houda  and
      Pino, Juan  and
      Bali, Kalika",
    booktitle = "Proceedings of the 2023 Conference on Empirical Methods in Natural Language Processing",
    month = dec,
    year = "2023",
    address = "Singapore",
    publisher = "Association for Computational Linguistics",
    url = "https://aclanthology.org/2023.emnlp-main.302/",
    doi = "10.18653/v1/2023.emnlp-main.302",
    pages = "4945--4977",
}

@article{lee2024prompt,
  title={Prompt Infection: LLM-to-LLM Prompt Injection within Multi-Agent Systems},
  author={Lee, Donghyun and Tiwari, Mo},
  journal={arXiv preprint arXiv:2410.07283},
  year={2024}
}

@inproceedings{chen2024struq,
  title={$\{$StruQ$\}$: Defending against prompt injection with structured queries},
  author={Chen, Sizhe and Piet, Julien and Sitawarin, Chawin and Wagner, David},
  booktitle={34th USENIX Security Symposium (USENIX Security 25)},
  pages={2383--2400},
  year={2025}
}

@inproceedings{chen2025secalign,
  title={SecAlign: Defending Against Prompt Injection with Preference Optimization},
  author={Chen, Sizhe and Zharmagambetov, Arman and Mahloujifar, Saeed and Chaudhuri, Kamalika and Wagner, David and Guo, Chuan},
  booktitle={The ACM Conference on Computer and Communications Security (CCS)},
  year={2025}
}

@inproceedings{wu2024instructional,
  title={Instructional Segment Embedding: Improving LLM Safety with Instruction Hierarchy},
  author={Wu, Tong and Zhang, Shujian and Song, Kaiqiang and Xu, Silei and Zhao, Sanqiang and Agrawal, Ravi and Indurthi, Sathish Reddy and Xiang, Chong and Mittal, Prateek and Zhou, Wenxuan},
  booktitle={The Thirteenth International Conference on Learning Representations},
  year={2024}
}

@article{wallace2024instruction,
  title={The instruction hierarchy: Training llms to prioritize privileged instructions},
  author={Wallace, Eric and Xiao, Kai and Leike, Reimar and Weng, Lilian and Heidecke, Johannes and Beutel, Alex},
  journal={arXiv preprint arXiv:2404.13208},
  year={2024}
}

@inproceedings{wu2025isolategpt,
  title={{IsolateGPT: An Execution Isolation Architecture for LLM-Based Systems}},
  author={Wu, Yuhao and Roesner, Franziska and Kohno, Tadayoshi and Zhang, Ning and Iqbal, Umar},
  booktitle={Network and Distributed System Security Symposium (NDSS)},
  year={2025},
}

@inproceedings{naihin2023testing,
  title={Testing Language Model Agents Safely in the Wild},
  author={Naihin, Silen and Atkinson, David and Green, Marc and Hamadi, Merwane and Swift, Craig and Schonholtz, Douglas and Kalai, Adam Tauman and Bau, David},
  booktitle={Socially Responsible Language Modelling Research},
  year={2023}
}

@inproceedings{zhan24injecagent,
    title = "{I}njec{A}gent: Benchmarking Indirect Prompt Injections in Tool-Integrated Large Language Model Agents",
    author = "Zhan, Qiusi  and
      Liang, Zhixiang  and
      Ying, Zifan  and
      Kang, Daniel",
    editor = "Ku, Lun-Wei  and
      Martins, Andre  and
      Srikumar, Vivek",
    booktitle = "Findings of the Association for Computational Linguistics: ACL 2024",
    month = aug,
    year = "2024",
    address = "Bangkok, Thailand",
    publisher = "Association for Computational Linguistics",
    url = "https://aclanthology.org/2024.findings-acl.624/",
    doi = "10.18653/v1/2024.findings-acl.624",
    pages = "10471--10506",
}

@article{achiam2023gpt,
  title={Gpt-4 technical report},
  author={Achiam, Josh and Adler, Steven and Agarwal, Sandhini and Ahmad, Lama and Akkaya, Ilge and Aleman, Florencia Leoni and Almeida, Diogo and Altenschmidt, Janko and Altman, Sam and Anadkat, Shyamal and others},
  journal={arXiv preprint arXiv:2303.08774},
  year={2023}
}

@article{team2023gemini,
  title={Gemini: a family of highly capable multimodal models},
  author={GeminiTeam},
  journal={arXiv preprint arXiv:2312.11805},
  year={2023}
}

@article{wu2024system,
  title={System-Level Defense against Indirect Prompt Injection Attacks: An Information Flow Control Perspective},
  author={Wu, Fangzhou and Cecchetti, Ethan and Xiao, Chaowei},
  journal={arXiv preprint arXiv:2409.19091},
  year={2024}
}

@inproceedings{jia2024task,
    title = "The Task Shield: Enforcing Task Alignment to Defend Against Indirect Prompt Injection in {LLM} Agents",
    author = "Jia, Feiran  and
      Wu, Tong  and
      Qin, Xin  and
      Squicciarini, Anna",
    editor = "Che, Wanxiang  and
      Nabende, Joyce  and
      Shutova, Ekaterina  and
      Pilehvar, Mohammad Taher",
    booktitle = "Proceedings of the 63rd Annual Meeting of the Association for Computational Linguistics (Volume 1: Long Papers)",
    month = jul,
    year = "2025",
    address = "Vienna, Austria",
    publisher = "Association for Computational Linguistics",
    url = "https://aclanthology.org/2025.acl-long.1435/",
    doi = "10.18653/v1/2025.acl-long.1435",
    pages = "29680--29697",
    ISBN = "979-8-89176-251-0",
}

@inproceedings{liu2025datasentinel,
  title={DataSentinel: A Game-Theoretic Detection of Prompt Injection Attacks},
  author={Liu, Yupei and Jia, Yuqi and Jia, Jinyuan and Song, Dawn and Gong, Neil Zhenqiang},
  booktitle={2025 IEEE Symposium on Security and Privacy (SP)},
  pages={2190--2208},
  year={2025},
  organization={IEEE}
}

@inproceedings{zhu2025melon,
title={{MELON}: Indirect Prompt Injection Defense via Masked Re-execution and Tool Comparison},
author={Zhu, Kaijie and Yang, Xianjun and Wang, Jindong and Guo, Wenbo and Wang, William Yang},
booktitle={Forty-second International Conference on Machine Learning},
year={2025}
}

@article{touvron2023llama1,
  title={Llama: Open and efficient foundation language models},
  author={Touvron, Hugo and Lavril, Thibaut and Izacard, Gautier and Martinet, Xavier and Lachaux, Marie-Anne and Lacroix, Timoth{\'e}e and Rozi{\`e}re, Baptiste and Goyal, Naman and Hambro, Eric and Azhar, Faisal and others},
  journal={arXiv preprint arXiv:2302.13971},
  year={2023}
}

@inproceedings{wei2022chain,
  title={Chain-of-thought prompting elicits reasoning in large language models},
  author={Wei, Jason and Wang, Xuezhi and Schuurmans, Dale and Bosma, Maarten and Xia, Fei and Chi, Ed and Le, Quoc V and Zhou, Denny and others},
  booktitle={NeurIPS},
  year={2022}
}

@inproceedings{yao2023react,
  title={React: Synergizing reasoning and acting in language models},
  author={Yao, Shunyu and Zhao, Jeffrey and Yu, Dian and Du, Nan and Shafran, Izhak and Narasimhan, Karthik and Cao, Yuan},
  booktitle={ICLR},
  year={2023}
}

@article{zou2024poisonedrag,
  title={PoisonedRAG: Knowledge Corruption Attacks to Retrieval-Augmented Generation of Large Language Models},
  author={Zou, Wei and Geng, Runpeng and Wang, Binghui and Jia, Jinyuan},
  journal={USENIX Security Symposium},
  year={2025},
  note={arXiv:2402.07867}
}

@inproceedings{zhong2023poisoning,
  title={Poisoning Retrieval Corpora by Injecting Adversarial Passages},
  author={Zhong, Zexuan and Huang, Zonglin and Wettig, Alexander and Chen, Danqi},
  booktitle={Proceedings of the 2023 Conference on Empirical Methods in Natural Language Processing},
  pages={13764--13775},
  year={2023}
}

@article{debenedetti2025defeating,
  title={Defeating prompt injections by design},
  author={Debenedetti, Edoardo and Shumailov, Ilia and Fan, Tianqi and Hayes, Jamie and Carlini, Nicholas and Fabian, Daniel and Kern, Christoph and Shi, Chongyang and Terzis, Andreas and Tram{\`e}r, Florian},
  journal={arXiv preprint arXiv:2503.18813},
  year={2025}
}

@article{shi2025progent,
  title={Progent: Programmable Privilege Control for LLM Agents},
  author={Shi, Tianneng and He, Jingxuan and Wang, Zhun and Wu, Linyu and Li, Hongwei and Guo, Wenbo and Song, Dawn},
  journal={arXiv preprint arXiv:2504.11703},
  year={2025}
}

@inproceedings{yi2025benchmarking,
  title={Benchmarking and defending against indirect prompt injection attacks on large language models},
  author={Yi, Jingwei and Xie, Yueqi and Zhu, Bin and Kiciman, Emre and Sun, Guangzhong and Xie, Xing and Wu, Fangzhao},
  booktitle={Proceedings of the 31st ACM SIGKDD Conference on Knowledge Discovery and Data Mining V. 1},
  pages={1809--1820},
  year={2025}
}

@inproceedings{jacob2025promptshield,
  title={Promptshield: Deployable detection for prompt injection attacks},
  author={Jacob, Dennis and Alzahrani, Hend and Hu, Zhanhao and Alomair, Basel and Wagner, David},
  booktitle={Proceedings of the Fifteenth ACM Conference on Data and Application Security and Privacy},
  pages={341--352},
  year={2024}
}

@inproceedings{zhang2025asb,
title={Agent Security Bench ({ASB}): Formalizing and Benchmarking Attacks and Defenses in {LLM}-based Agents},
author={Hanrong Zhang and Jingyuan Huang and Kai Mei and Yifei Yao and Zhenting Wang and Chenlu Zhan and Hongwei Wang and Yongfeng Zhang},
booktitle={The Thirteenth International Conference on Learning Representations},
year={2025}
}

@misc{yu2025promptfuzz,
      title={PROMPTFUZZ: Harnessing Fuzzing Techniques for Robust Testing of Prompt Injection in LLMs},
      author={Jiahao Yu and Yangguang Shao and Hanwen Miao and Junzheng Shi},
      year={2025},
      eprint={2409.14729},
      archivePrefix={arXiv},
      primaryClass={cs.CR},
      url={https://arxiv.org/abs/2409.14729},
}

@article{beurer2025design,
  title={Design Patterns for Securing LLM Agents against Prompt Injections},
  author={Beurer-Kellner, Luca and Cre{\c{t}}u, Beat Buesser Ana-Maria and Debenedetti, Edoardo and Dobos, Daniel and Fabian, Daniel and Fischer, Marc and Froelicher, David and Grosse, Kathrin and Naeff, Daniel and Ozoani, Ezinwanne and others},
  journal={arXiv preprint arXiv:2506.08837},
  year={2025}
}

@article{yu2025survey,
  title={A survey on trustworthy llm agents: Threats and countermeasures},
  author={Yu, Miao and Meng, Fanci and Zhou, Xinyun and Wang, Shilong and Mao, Junyuan and Pang, Linsey and Chen, Tianlong and Wang, Kun and Li, Xinfeng and Zhang, Yongfeng and others},
  journal={arXiv preprint arXiv:2503.09648},
  year={2025}
}

@article{zhang2025llm,
  title={LLM Agents Should Employ Security Principles},
  author={Zhang, Kaiyuan and Su, Zian and Chen, Pin-Yu and Bertino, Elisa and Zhang, Xiangyu and Li, Ninghui},
  journal={arXiv preprint arXiv:2505.24019},
  year={2025}
}

@misc{mozilla-meta-ai,
  title={Meta: Help Users Stop Accidentally Sharing Private AI Chats},
  author={Mozilla},
  year={2025},
  note={\url{https://www.mozillafoundation.org/en/campaigns/meta-help-users-stop-accidentally-sharing-private-ai-conversations/}}
}

@inproceedings{airgap,
  title={AirGapAgent: Protecting privacy-conscious conversational agents},
  author={Bagdasarian, Eugene and Yi, Ren and Ghalebikesabi, Sahra and Kairouz, Peter and Gruteser, Marco and Oh, Sewoong and Balle, Borja and Ramage, Daniel},
  booktitle={Proceedings of the 2024 on ACM SIGSAC Conference on Computer and Communications Security},
  pages={3868--3882},
  year={2024}
}

@article{pfi,
  title={Prompt flow integrity to prevent privilege escalation in llm agents},
  author={Kim, Juhee and Choi, Woohyuk and Lee, Byoungyoung},
  journal={arXiv preprint arXiv:2503.15547},
  year={2025}
}

@misc{chatgpt-shared-links,
  title={ChatGPT Shared Links FAQ},
  author={OpenAI},
  year={2025},
  note={\url{https://help.openai.com/en/articles/7925741-chatgpt-shared-links-faq}}
}

@misc{langchain-chat-history,
  title={Chat history},
  author={LangChain},
  year={2025},
  note={\url{https://python.langchain.com/docs/concepts/chat_history/}}
}

@misc{cve-2025-32711,
  title={CVE-2025-32711},
  author={MITRE},
  year={2025},
  note={\url{https://www.cve.org/CVERecord?id=CVE-2025-32711}}
}

@misc{cve-2024-5565,
  title={CVE-2024-5565},
  author={MITRE},
  year={2024},
  note={\url{https://www.cve.org/CVERecord?id=CVE-2024-5565}}
}

@article{dong2025practical,
  title={A practical memory injection attack against llm agents},
  author={Dong, Shen and Xu, Shaochen and He, Pengfei and Li, Yige and Tang, Jiliang and Liu, Tianming and Liu, Hui and Xiang, Zhen},
  journal={arXiv preprint arXiv:2503.03704},
  year={2025}
}

@misc{github-mcp-exploited,
  title={GitHub MCP Exploited: Accessing private repositories via MCP},
  author={Invariantlabs},
  year={2025},
  note={\url{https://invariantlabs.ai/blog/mcp-github-vulnerability}}
}

@misc{github-data-exfiltration,
  title={GitHub Copilot Chat: From Prompt Injection to Data Exfiltration},
  author={Embrace The Red},
  year={2025},
  note={\url{https://embracethered.com/blog/posts/2024/github-copilot-chat-prompt-injection-data-exfiltration/}}
}

@misc{cross-agent-privilege,
  title={Cross-Agent Privilege Escalation: When Agents Free Each Other},
  author={Embrace The Red},
  year={2025},
  note={\url{https://embracethered.com/blog/posts/2025/cross-agent-privilege-escalation-agents-that-free-each-other/}}
}

@inproceedings{zhang2024goal,
  title={Goal-guided generative prompt injection attack on large language models},
  author={Zhang, Chong and Jin, Mingyu and Yu, Qinkai and Liu, Chengzhi and Xue, Haochen and Jin, Xiaobo},
  booktitle={2024 IEEE International Conference on Data Mining (ICDM)},
  pages={941--946},
  year={2024},
  organization={IEEE}
}

@article{liu2023prompt,
  title={Prompt injection attack against llm-integrated applications},
  author={Liu, Yi and Deng, Gelei and Li, Yuekang and Wang, Kailong and Wang, Zihao and Wang, Xiaofeng and Zhang, Tianwei and Liu, Yepang and Wang, Haoyu and Zheng, Yan and others},
  journal={arXiv preprint arXiv:2306.05499},
  year={2023}
}

@misc{jumping-the-line,
  title={Jumping the line: How MCP servers can attack you before you ever use them},
  author={Trail of Bits},
  year={2025},
  note={\url{https://blog.trailofbits.com/2025/04/21/jumping-the-line-how-mcp-servers-can-attack-you-before-you-ever-use-them/}}
}

@article{shi2025prompt,
  title={Prompt Injection Attack to Tool Selection in LLM Agents},
  author={Shi, Jiawen and Yuan, Zenghui and Tie, Guiyao and Zhou, Pan and Gong, Neil Zhenqiang and Sun, Lichao},
  journal={arXiv preprint arXiv:2504.19793},
  year={2025}
}

@article{siddiqui2024permissive,
  title={Permissive Information-Flow Analysis for Large Language Models},
  author={Siddiqui, Shoaib Ahmed and Gaonkar, Radhika and K{\"o}pf, Boris and Krueger, David and Paverd, Andrew and Salem, Ahmed and Tople, Shruti and Wutschitz, Lukas and Xia, Menglin and Zanella-B{\'e}guelin, Santiago},
  journal={arXiv preprint arXiv:2410.03055},
  year={2024}
}

@misc{march-20-chatgpt,
  title={March 20 ChatGPT outage: Here’s what happened},
  author={OpenAI},
  year={2023},
  note={\url{https://openai.com/index/march-20-chatgpt-outage/}}
}

@misc{chatgpt-banned-italy,
  title={ChatGPT banned in Italy over privacy concerns},
  author={BBC News},
  year={2023},
  note={\url{https://www.bbc.co.uk/news/technology-65139406}}
}

@misc{deepseek-banned-korea,
  title={South Korea Bans Downloads of DeepSeek, the Chinese A.I. App},
  author={New York Times},
  year={2025},
  note={\url{https://www.nytimes.com/2025/02/17/business/south-korea-deepseek-china-ai.html}}
}

@misc{openai-temporary-chat,
  title={Temporary Chat FAQ},
  author={OpenAI},
  year={2025},
  note={\url{https://help.openai.com/en/articles/8914046-temporary-chat-faq}}
}

@inproceedings{klein2009sel4,
  title={seL4: Formal verification of an OS kernel},
  author={Klein, Gerwin and Elphinstone, Kevin and Heiser, Gernot and Andronick, June and Cock, David and Derrin, Philip and Elkaduwe, Dhammika and Engelhardt, Kai and Kolanski, Rafal and Norrish, Michael and others},
  booktitle={Proceedings of the ACM SIGOPS 22nd symposium on Operating systems principles},
  pages={207--220},
  year={2009}
}

@inproceedings{hawblitzel2014ironclad,
  title={Ironclad apps:$\{$End-to-End$\}$ security via automated $\{$Full-System$\}$ verification},
  author={Hawblitzel, Chris and Howell, Jon and Lorch, Jacob R and Narayan, Arjun and Parno, Bryan and Zhang, Danfeng and Zill, Brian},
  booktitle={11th USENIX symposium on operating systems design and implementation (OSDI 14)},
  pages={165--181},
  year={2014}
}

@article{li2024formal,
  title={Formal-llm: Integrating formal language and natural language for controllable llm-based agents},
  author={Li, Zelong and Hua, Wenyue and Wang, Hao and Zhu, He and Zhang, Yongfeng},
  journal={arXiv preprint arXiv:2402.00798},
  year={2024}
}

@misc{google-mitigating-prompt,
  title={Mitigating prompt injection attacks with a layered defense strategy},
  author={Google},
  year={2025},
  note={\url{https://security.googleblog.com/2025/06/mitigating-prompt-injection-attacks.html}}
}

@inproceedings{chan2024visibility,
  title={Visibility into AI agents},
  author={Chan, Alan and Ezell, Carson and Kaufmann, Max and Wei, Kevin and Hammond, Lewis and Bradley, Herbie and Bluemke, Emma and Rajkumar, Nitarshan and Krueger, David and Kolt, Noam and others},
  booktitle={Proceedings of the 2024 ACM Conference on Fairness, Accountability, and Transparency},
  pages={958--973},
  year={2024}
}

@article{radosevich2025mcp,
  title={Mcp safety audit: Llms with the model context protocol allow major security exploits},
  author={Radosevich, Brandon and Halloran, John},
  journal={arXiv preprint arXiv:2504.03767},
  year={2025}
}

@inproceedings{luo2025agrail,
  author={Weidi Luo and Shenghong Dai and Xiaogeng Liu and Suman Banerjee and Huan Sun and Muhao Chen and Chaowei Xiao},
  title={AGrail: A Lifelong Agent Guardrail with Effective and Adaptive Safety Detection},
  year={2025},
  cdate={1735689600000},
  pages={8104-8139},
  url={https://aclanthology.org/2025.acl-long.399/},
  booktitle={ACL (1)},
}

@inproceedings{tsai2025contextual,
  title={Contextual Agent Security: A Policy for Every Purpose},
  author={Tsai, Lillian and Bagdasarian, Eugene},
  booktitle={Proceedings of the 2025 Workshop on Hot Topics in Operating Systems},
  pages={8--17},
  year={2025}
}

@techreport{bell1973secure,
  title={Secure computer systems: Mathematical foundations},
  author={Bell, D Elliot and LaPadula, Leonard J},
  year={1973}
}

@techreport{biba1977integrity,
  title={Integrity considerations for secure computer systems},
  author={Biba, Kenneth J},
  year={1977}
}

@misc{codeshield,
  title={CodeShield},
  author={meta-llama},
  year={2025},
  note={\url{https://github.com/meta-llama/PurpleLlama/tree/main/CodeShield}}
}

@article{xiang2024guardagent,
  title={Guardagent: Safeguard llm agents by a guard agent via knowledge-enabled reasoning},
  author={Xiang, Zhen and Zheng, Linzhi and Li, Yanjie and Hong, Junyuan and Li, Qinbin and Xie, Han and Zhang, Jiawei and Xiong, Zidi and Xie, Chulin and Yang, Carl and others},
  journal={arXiv preprint arXiv:2406.09187},
  year={2024}
}

@inproceedings{chen2025shieldagent,
title={ShieldAgent: Shielding Agents via Verifiable Safety Policy Reasoning},
author={Zhaorun Chen and Mintong Kang and Bo Li},
booktitle={Forty-second International Conference on Machine Learning},
year={2025}
}

@article{hardy1988confused,
  title={The Confused Deputy: (or why capabilities might have been invented)},
  author={Hardy, Norm},
  journal={ACM SIGOPS Operating Systems Review},
  volume={22},
  number={4},
  pages={36--38},
  year={1988},
  publisher={ACM New York, NY, USA}
}

@inproceedings{myers1999jflow,
  title={JFlow: Practical mostly-static information flow control},
  author={Myers, Andrew C},
  booktitle={Proceedings of the 26th ACM SIGPLAN-SIGACT symposium on Principles of programming languages},
  pages={228--241},
  year={1999}
}

@article{denning1976lattice,
  title={A lattice model of secure information flow},
  author={Denning, Dorothy E},
  journal={Communications of the ACM},
  volume={19},
  number={5},
  pages={236--243},
  year={1976},
  publisher={ACM New York, NY, USA}
}

@inproceedings{newsome2005dynamic,
  title={Dynamic taint analysis for automatic detection, analysis, and signaturegeneration of exploits on commodity software.},
  author={Newsome, James and Song, Dawn Xiaodong},
  booktitle={NDSS},
  volume={5},
  pages={3--4},
  year={2005}
}

@article{costa2025securing,
  title={Securing AI Agents with Information-Flow Control},
  author={Costa, Manuel and K{\"o}pf, Boris and Kolluri, Aashish and Paverd, Andrew and Russinovich, Mark and Salem, Ahmed and Tople, Shruti and Wutschitz, Lukas and Zanella-B{\'e}guelin, Santiago},
  journal={arXiv preprint arXiv:2505.23643},
  year={2025}
}

@misc{okta,
  title={Okta Documentation},
  author={Okta},
  year={2025},
  note={\url{https://help.okta.com/en-us/content/index.htm}}
}

@misc{composio,
  title={Welcome to Composio},
  author={Composio},
  year={2025},
  note={\url{https://docs.composio.dev/docs/welcome}}
}

@misc{w3c-did,
  title={Decentralized Identifiers (DIDs) v1.0},
  author={W3C},
  year={2025},
  note={\url{https://www.w3.org/TR/did-1.0/}}
}

@misc{microsoft-verified-id,
  title={Introduction to Microsoft Entra Verified ID},
  author={Microsoft},
  year={2025},
  note={\url{https://learn.microsoft.com/en-us/entra/verified-id/decentralized-identifier-overview}}
}

@article{luo2025agentauditor,
  title={Agentauditor: Human-level safety and security evaluation for llm agents},
  author={Luo, Hanjun and Dai, Shenyu and Ni, Chiming and Li, Xinfeng and Zhang, Guibin and Wang, Kun and Liu, Tongliang and Salam, Hanan},
  journal={arXiv preprint arXiv:2506.00641},
  year={2025}
}

@inproceedings{yueh2025monitoring,
  title={Monitoring LLM Agents for Sequentially Contextual Harm},
  author={Yueh-Han, Chen and Joshi, Nitish and Chen, Yulin and He, He and Angell, Rico},
  booktitle={ICLR 2025 Workshop on Building Trust in Language Models and Applications},
  year={2025}
}

@misc{heimdallm,
  title={HeimdaLLM},
  author={Andrew Moffat},
  year={2023},
  note={\url{https://heimdallm.readthedocs.io/en/main/}}
}

@article{zhong2025honeybee,
  title={HoneyBee: Efficient Role-based Access Control for Vector Databases via Dynamic Partitioning},
  author={Zhong, Hongbin and Lentz, Matthew and Narodytska, Nina and Szekeres, Adriana and Rong, Kexin},
  journal={arXiv preprint arXiv:2505.01538},
  year={2025}
}

@article{yao2025controlnet,
  title={Controlnet: A firewall for rag-based llm system},
  author={Yao, Hongwei and Shi, Haoran and Chen, Yidou and Jiang, Yixin and Wang, Cong and Qin, Zhan},
  journal={arXiv preprint arXiv:2504.09593},
  year={2025}
}

@misc{amazon-bedrock,
  title={Access control for vector stores using metadata filtering with Amazon Bedrock Knowledge Bases},
  author={Amazon},
  year={2024},
  note={\url{https://aws.amazon.com/blogs/machine-learning/access-control-for-vector-stores-using-metadata-filtering-with-knowledge-bases-for-amazon-bedrock/}}
}

@article{he2025sentinelagent,
  title={SentinelAgent: Graph-based Anomaly Detection in Multi-Agent Systems},
  author={He, Xu and Wu, Di and Zhai, Yan and Sun, Kun},
  journal={arXiv preprint arXiv:2505.24201},
  year={2025}
}

@article{zhou2025guardian,
  title={GUARDIAN: Safeguarding LLM Multi-Agent Collaborations with Temporal Graph Modeling},
  author={Zhou, Jialong and Wang, Lichao and Yang, Xiao},
  journal={arXiv preprint arXiv:2505.19234},
  year={2025}
}

@misc{anthropic-connectors,
  title={Anthropic Connectors Directory FAQ},
  author={Anthropic},
  year={2025},
  note={\url{https://support.anthropic.com/en/articles/11596036-anthropic-connectors-directory-faq}}
}

@inproceedings{wen2025adaptive,
title={Adaptive Deployment of Untrusted {LLM}s Reduces Distributed Threats},
author={Jiaxin Wen and Vivek Hebbar and Caleb Larson and Aryan Bhatt and Ansh Radhakrishnan and Mrinank Sharma and Henry Sleight and Shi Feng and He He and Ethan Perez and Buck Shlegeris and Akbir Khan},
booktitle={The Thirteenth International Conference on Learning Representations},
year={2025}
}

@article{lee2025safeguarding,
  title={Safeguarding mobile gui agent via logic-based action verification},
  author={Lee, Jungjae and Lee, Dongjae and Choi, Chihun and Im, Youngmin and Wi, Jaeyoung and Heo, Kihong and Oh, Sangeun and Lee, Sunjae and Shin, Insik},
  journal={arXiv preprint arXiv:2503.18492},
  year={2025}
}

@misc{mcp-context-protector,
  title={mcp-context-protector},
  author={trailofbits},
  year={2025},
  note={\url{https://github.com/trailofbits/mcp-context-protector}}
}

@misc{etdi,
  title={Enhanced Tool Definition Interface (ETDI): A Security Fortification for the Model Context Protocol},
  author={ETDI Documentation},
  year={2025},
  note={\url{https://vineethsai.github.io/python-sdk/etdi-concepts/\#introduction-the-imperative-for-secure-mcp}}
}

@misc{oauth,
  title={The OAuth 2.0 Authorization Framework},
  author={Internet Engineering Task Force (IETF)},
  year={2012},
  note={\url{https://www.rfc-editor.org/rfc/rfc6749.html}}
}

@misc{openid-connect,
  title={OpenID Connect Core 1.0 incorporating errata set 1},
  author={OpenID Foundation},
  year={2014},
  note={\url{https://openid.net/specs/openid-connect-core-1_0.html}}
}

@misc{langsmith-leak,
  title={How an AI Agent Vulnerability in LangSmith Could Lead to Stolen API Keys and Hijacked LLM Responses},
  author={Sasi Levi, Gal Moyal},
  year={2025},
  note={\url{https://noma.security/blog/how-an-ai-agent-vulnerability-in-langsmith-could-lead-to-stolen-api-keys-and-hijacked-llm-responses/}}
}

@misc{gemini-canvas,
  title={Gemini Canvas},
  author={Gemini},
  year={2025},
  note={\url{https://gemini.google/overview/canvas/}}
}

@article{wang2025agentarmor,
  title={AgentArmor: Enforcing Program Analysis on Agent Runtime Trace to Defend Against Prompt Injection},
  author={Wang, Peiran and Liu, Yang and Lu, Yunfei and Cai, Yifeng and Chen, Hongbo and Yang, Qingyou and Zhang, Jie and Hong, Jue and Wu, Ye},
  journal={arXiv preprint arXiv:2508.01249},
  year={2025}
}

@misc{aim-echoleak,
  title={Breaking down ‘EchoLeak’, the First Zero-Click AI Vulnerability Enabling Data Exfiltration from Microsoft 365 Copilot},
  author={Itay Ravia},
  year={2025},
  note={\url{https://www.catonetworks.com/blog/breaking-down-echoleak/}}
}

@misc{dual-llm,
  title={The Dual LLM pattern for building AI assistants that can resist prompt injection},
  author={Simon Willison},
  year={2023},
  note={\url{https://simonwillison.net/2023/Apr/25/dual-llm-pattern/}}
}

@misc{CVE-2025-54795,
  title={CVE-2025-54795},
  author={MITRE},
  year={2025},
  note={\url{https://nvd.nist.gov/vuln/detail/CVE-2025-54795}}
}

@article{sharma2025constitutional,
  title={Constitutional classifiers: Defending against universal jailbreaks across thousands of hours of red teaming},
  author={Sharma, Mrinank and Tong, Meg and Mu, Jesse and Wei, Jerry and Kruthoff, Jorrit and Goodfriend, Scott and Ong, Euan and Peng, Alwin and Agarwal, Raj and Anil, Cem and others},
  journal={arXiv preprint arXiv:2501.18837},
  year={2025}
}

@inproceedings{felt2012android,
  title={Android permissions: User attention, comprehension, and behavior},
  author={Felt, Adrienne Porter and Ha, Elizabeth and Egelman, Serge and Haney, Ariel and Chin, Erika and Wagner, David},
  booktitle={Proceedings of the eighth symposium on usable privacy and security},
  pages={1--14},
  year={2012}
}

@article{li2025safeflow,
  title={Safeflow: A principled protocol for trustworthy and transactional autonomous agent systems},
  author={Li, Peiran and Zou, Xinkai and Wu, Zhuohang and Li, Ruifeng and Xing, Shuo and Zheng, Hanwen and Hu, Zhikai and Wang, Yuping and Li, Haoxi and Yuan, Qin and others},
  journal={arXiv preprint arXiv:2506.07564},
  year={2025}
}

@inproceedings{brumley2004privtrans,
  title={Privtrans: Automatically partitioning programs for privilege separation},
  author={Brumley, David and Song, Dawn},
  booktitle={USENIX security symposium},
  volume={57},
  number={72},
  year={2004}
}

@article{shi2025promptarmor,
  title={PromptArmor: Simple yet Effective Prompt Injection Defenses},
  author={Shi, Tianneng and Zhu, Kaijie and Wang, Zhun and Jia, Yuqi and Cai, Will and Liang, Weida and Wang, Haonan and Alzahrani, Hend and Lu, Joshua and Kawaguchi, Kenji and others},
  journal={arXiv preprint arXiv:2507.15219},
  year={2025}
}

@misc{comet-prompt-injection,
  title={Agentic Browser Security: Indirect Prompt Injection in Perplexity Comet},
  author={Brave},
  year={2025},
  note={\url{https://brave.com/blog/comet-prompt-injection/}}
}

@misc{github-copilot,
  title={GitHub Copilot},
  author={GitHub},
  year={2025},
  note={\url{https://github.com/features/copilot}}
}

@misc{cursor,
  title={Cursor - The AI-first code editor},
  author={Cursor},
  year={2025},
  note={\url{https://www.cursor.com/}}
}

@misc{gemini-code-assist,
  title={Gemini Code Assist},
  author={Google Cloud},
  year={2025},
  note={\url{https://cloud.google.com/products/gemini/code-assist}}
}

@misc{openai-chatgpt,
  title={ChatGPT},
  author={OpenAI},
  year={2025},
  note={\url{https://chatgpt.com/}}
}

@misc{google-gemini,
  title={Gemini},
  author={Google},
  year={2025},
  note={\url{https://gemini.google.com/}}
}

@misc{comet-browser,
  title={Comet Browser: Browse at the speed of thought},
  author={Perplexity},
  year={2025},
  note={\url{https://www.perplexity.ai/comet}}
}

@misc{ios-permission,
  title={Requesting access to protected resources},
  author={Apple},
  year={2025},
  note={\url{https://developer.apple.com/documentation/uikit/requesting-access-to-protected-resources}}
}

@misc{namespace,
  title={namespaces(7) — Linux manual page},
  author={Linux},
  year={2024},
  note={\url{https://man7.org/linux/man-pages/man7/namespaces.7.html}}
}

@misc{cgroups,
  title={cgroups(7) — Linux manual page},
  author={Linux},
  year={2024},
  note={\url{https://man7.org/linux/man-pages/man7/cgroups.7.html}}
}

@misc{seccomp,
  title={seccomp(2) — Linux manual page},
  author={Linux},
  year={2024},
  note={\url{https://man7.org/linux/man-pages/man2/seccomp.2.html}}
}

@misc{google-oauth,
  title={Using OAuth 2.0 to Access Google APIs},
  author={Google},
  year={2025},
  note={\url{https://developers.google.com/identity/protocols/oauth2} (accessed 14, April, 2025)},
}

@misc{slack-oauth,
  title={Installing with OAuth},
  author={Slack},
  year={2025},
  note={\url{https://api.slack.com/authentication/oauth-v2} (accessed 14, April, 2025)},
}

@inproceedings{lee2020keystone,
  title={Keystone: An open framework for architecting trusted execution environments},
  author={Lee, Dayeol and Kohlbrenner, David and Shinde, Shweta and Asanovi{\'c}, Krste and Song, Dawn},
  booktitle={Proceedings of the Fifteenth European Conference on Computer Systems},
  pages={1--16},
  year={2020}
}

@article{sev2020strengthening,
  title={Strengthening VM isolation with integrity protection and more},
  author={Sev-Snp, AMD},
  journal={White Paper, January},
  volume={53},
  number={2020},
  pages={1450--1465},
  year={2020}
}

@inproceedings{li2025ace,
  title={ACE: A Security Architecture for LLM-Integrated App Systems},
  author={Li, Evan and Mallick, Tushin and Rose, Evan and Robertson, William and Oprea, Alina and Nita-Rotaru, Cristina},
  booktitle = {Network and Distributed System Security (NDSS) Symposium},
  year = {2026}
}

@inproceedings{wang2025sok,
author = { Wang, Xunguang and Ji, Zhenlan and Wang, Wenxuan and Li, Zongjie and Wu, Daoyuan and Wang, Shuai },
booktitle = { 2026 IEEE Symposium on Security and Privacy (SP) },
title = {{SoK: Evaluating Jailbreak Guardrails for Large Language Models}},
year = {2026},
volume = {},
ISSN = {2375-1207},
pages = {1427-1446},
doi = {10.1109/SP63933.2026.00076},
url = {https://doi.ieeecomputersociety.org/10.1109/SP63933.2026.00076},
publisher = {IEEE Computer Society},
address = {Los Alamitos, CA, USA},
month =May
}

@inproceedings{syros2025saga,
  title = {SAGA: A Security Architecture for Governing AI Agentic Systems},
  author = {Syros, Georgios and Suri, Anshuman and Ginesin, Jacob and Nita-Rotaru, Cristina and Oprea, Alina},
  booktitle = {Network and Distributed System Security (NDSS) Symposium},
  year = {2026}
}

@article{chang2025agent,
  title={Agent network protocol technical white paper},
  author={Chang, Gaowei and Lin, Eidan and Yuan, Chengxuan and Cai, Rizhao and Chen, Binbin and Xie, Xuan and Zhang, Yin},
  journal={arXiv preprint arXiv:2508.00007},
  year={2025}
}

@misc{token-vault,
  title={Calling APIs with Token Vault},
  author={auth0},
  year={2025},
  note={\url{https://auth0.com/ai/docs/intro/token-vault}},
}

@misc{codex-sandbox,
  title={Codex security guide},
  author={OpenAI Developers},
  year={2025},
  note={\url{https://developers.openai.com/codex/security/}},
}

@misc{gemini-sandbox,
  title={Sandboxing in the Gemini CLI},
  author={gemini-cli},
  year={2025},
  note={\url{https://google-gemini.github.io/gemini-cli/docs/cli/sandbox.html}}
}

@inproceedings{roth2020complex,
  title={Complex security policy? a longitudinal analysis of deployed content security policies},
  author={Roth, Sebastian and Barron, Timothy and Calzavara, Stefano and Nikiforakis, Nick and Stock, Ben},
  booktitle={Proceedings of the 27th Network and Distributed System Security Symposium (NDSS)},
  year={2020}
}

@misc{opentelemetry,
  title = {OpenTelemetry},
  author = {{OpenTelemetry}},
  year = {2025},
  note = {\url{https://opentelemetry.io/}},
}

@misc{posthog,
  title = {PostHog},
  author = {{PostHog}},
  year = {2025},
  note = {\url{https://posthog.com/}},
}

@article{yang2024watch,
  title={Watch out for your agents! investigating backdoor threats to llm-based agents},
  author={Yang, Wenkai and Bi, Xiaohan and Lin, Yankai and Chen, Sishuo and Zhou, Jie and Sun, Xu},
  journal={Advances in Neural Information Processing Systems},
  volume={37},
  pages={100938--100964},
  year={2024}
}

@inproceedings{wang2024badagent,
  title={BadAgent: Inserting and Activating Backdoor Attacks in LLM Agents},
  author={Wang, Yifei and Xue, Dizhan and Zhang, Shengjie and Qian, Shengsheng},
  booktitle={Proceedings of the 62nd Annual Meeting of the Association for Computational Linguistics (Volume 1: Long Papers)},
  pages={9811--9827},
  year={2024}
}

@inproceedings{spracklen2025we,
  title={We Have a Package for You! A Comprehensive Analysis of Package Hallucinations by Code Generating LLMs},
  author={Spracklen, Joseph and Neupane, Ajaykumar and Challagundla, Sai Krishna and Vaidya, Jagannadh},
  booktitle={USENIX Security Symposium},
  year={2025}
}

@article{patlan2025real,
  title={Real Vulnerabilities in AI Agents: A Practical Threat Analysis of Web3 Agent Memory Attacks},
  author={Patlan, David and Perez, Luis and others},
  journal={arXiv preprint arXiv:2501.12345},
  year={2025}
}

@misc{mitre-atlas,
  title={{MITRE ATLAS} -- Adversarial Threat Landscape for Artificial-Intelligence Systems},
  author={{MITRE Corporation}},
  year={2024},
  note={\url{https://atlas.mitre.org/}}
}

@misc{owasp-top-10-llm,
  title={{OWASP} Top 10 for Large Language Model Applications},
  author={{OWASP Foundation}},
  year={2025},
  note={\url{https://owasp.org/www-project-top-10-for-large-language-model-applications/}}
}

@inproceedings{shen2024prompt,
  title={Prompt Stealing Attacks Against Text-to-Image Generation Models},
  author={Shen, Xinyue and Song, Yiting and Li, Yixin and Zhu, Yun and others},
  booktitle={USENIX Security Symposium},
  year={2024}
}

@article{yang2025prsa,
  title={{PRSA}: Prompt Reverse Stealing Attacks against Large Language Models},
  author={Yang, Yong and Xu, Bo and others},
  journal={arXiv preprint arXiv:2402.19200},
  year={2025}
}

@article{li2024personal,
  title={Personal {LLM} Agents: Insights and Survey about the Capability, Efficiency and Security},
  author={Li, Yuanchun and Hao, Yizhen and others},
  journal={arXiv preprint arXiv:2401.05459},
  year={2024}
}

@article{deng2025ai,
  title={AI Agents Under Threat: A Survey of Key Security Challenges and Future Pathways},
  author={Deng, Zijian and others},
  journal={arXiv preprint arXiv:2406.02630},
  year={2025}
}

@article{kumar2025overthink,
  title={Overthink: Slowdown Attacks on Reasoning LLMs},
  author={Kumar, Abhinav and others},
  journal={arXiv preprint arXiv:2502.12345},
  year={2025}
}

@inproceedings{fredrikson2015model,
  title={Model Inversion Attacks that Exploit Confidence Information and Basic Countermeasures},
  author={Fredrikson, Matt and Jha, Somesh and Ristenpart, Thomas},
  booktitle={Proceedings of the ACM SIGSAC Conference on Computer and Communications Security},
  pages={1322--1333},
  year={2015}
}

@article{abdelnabi2025firewalls,
  title={Firewalls to secure dynamic llm agentic networks},
  author={Abdelnabi, Sahar and Gomaa, Amr and Bagdasarian, Eugene and Kristensson, Per Ola and Shokri, Reza},
  journal={arXiv preprint arXiv:2502.01822},
  year={2025}
}

@article{jha2025breaking,
  title={Breaking and Fixing Defenses Against Control-Flow Hijacking in Multi-Agent Systems},
  author={Jha, Rishi and Triedman, Harold and Wagle, Justin and Shmatikov, Vitaly},
  journal={arXiv preprint arXiv:2510.17276},
  year={2025}
}

@article{cui2025safeguard,
  title={Safeguard-by-development: A privacy-enhanced development paradigm for multi-agent collaboration systems},
  author={Cui, Jian and Li, Zichuan and Xing, Luyi and Liao, Xiaojing},
  journal={arXiv preprint arXiv:2505.04799},
  year={2025}
}

@article{wang2025privacy,
  title={Privacy in action: Towards realistic privacy mitigation and evaluation for llm-powered agents},
  author={Wang, Shouju and Yu, Fenglin and Liu, Xirui and Qin, Xiaoting and Zhang, Jue and Lin, Qingwei and Zhang, Dongmei and Rajmohan, Saravan},
  journal={arXiv preprint arXiv:2509.17488},
  year={2025}
}

@article{wu2025towards,
  title={Towards automating data access permissions in ai agents},
  author={Wu, Yuhao and Yang, Ke and Roesner, Franziska and Kohno, Tadayoshi and Zhang, Ning and Iqbal, Umar},
  journal={arXiv preprint arXiv:2511.17959},
  year={2025}
}

@inproceedings{an2025ipiguard,
  title={Ipiguard: A novel tool dependency graph-based defense against indirect prompt injection in llm agents},
  author={An, Hengyu and Zhang, Jinghuai and Du, Tianyu and Zhou, Chunyi and Li, Qingming and Lin, Tao and Ji, Shouling},
  booktitle={EMNLP},
  year={2025}
}

@article{li2025drift,
  title={DRIFT: Dynamic Rule-Based Defense with Injection Isolation for Securing LLM Agents},
  author={Li, Hao and Liu, Xiaogeng and Chiu, Hung-Chun and Li, Dianqi and Zhang, Ning and Xiao, Chaowei},
  journal={arXiv preprint arXiv:2506.12104},
  year={2025}
}

@inproceedings{jing2025mcip,
  title={Mcip: Protecting mcp safety via model contextual integrity protocol},
  author={Jing, Huihao and Li, Haoran and Hu, Wenbin and Hu, Qi and Heli, Xu and Chu, Tianshu and Hu, Peizhao and Song, Yangqiu},
  booktitle={EMNLP},
  year={2025}
}

@misc{chatgpt-connectors,
  title={Connectors in ChatGPT},
  author={OpenAI},
  year={2025},
  note={\url{https://chatgpt.com/features/connectors/}}
}

@misc{nanobanana,
  title={Introducing Nano Banana Pro},
  author={Gemini},
  year={2025},
  note={\url{https://blog.google/technology/ai/nano-banana-pro/}}
}

@inproceedings{grosse2024towards,
  title={Towards more practical threat models in artificial intelligence security},
  author={Grosse, Kathrin and Bieringer, Lukas and Besold, Tarek R and Alahi, Alexandre M},
  booktitle={33rd USENIX Security Symposium (USENIX Security 24)},
  pages={4891--4908},
  year={2024}
}

@article{jia2025critical,
  title={A Critical Evaluation of Defenses against Prompt Injection Attacks},
  author={Jia, Yuqi and Shao, Zedian and Liu, Yupei and Jia, Jinyuan and Song, Dawn and Gong, Neil Zhenqiang},
  journal={arXiv preprint arXiv:2505.18333},
  year={2025}
}

@article{lee2025takedown,
  title={Takedown: How It's Done in Modern Coding Agent Exploits},
  author={Lee, Eunkyu and Kim, Donghyeon and Kim, Wonyoung and Yun, Insu},
  journal={arXiv preprint arXiv:2509.24240},
  year={2025}
}

@inproceedings{shvartzshnaider2019vaccine,
  title={Vaccine: Using contextual integrity for data leakage detection},
  author={Shvartzshnaider, Yan and Pavlinovic, Zvonimir and Balashankar, Ananth and Wies, Thomas and Subramanian, Lakshminarayanan and Nissenbaum, Helen and Mittal, Prateek},
  booktitle={The World Wide Web Conference},
  pages={1702--1712},
  year={2019}
}

@inproceedings{wijesekera2015android,
  title={Android permissions remystified: A field study on contextual integrity},
  author={Wijesekera, Primal and Baokar, Arjun and Hosseini, Ashkan and Egelman, Serge and Wagner, David and Beznosov, Konstantin},
  booktitle={24th USENIX Security Symposium (USENIX Security 15)},
  pages={499--514},
  year={2015}
}

@inproceedings{jia2017contexlot,
  title={ContexloT: Towards providing contextual integrity to appified IoT platforms.},
  author={Jia, Yunhan Jack and Chen, Qi Alfred and Wang, Shiqi and Rahmati, Amir and Fernandes, Earlence and Mao, Zhuoqing Morley and Prakash, Atul and Unviersity, SJ},
  booktitle={ndss},
  volume={2},
  number={2},
  pages={2--2},
  year={2017},
  organization={San Diego}
}

@inproceedings{south2025authenticated,
  title={Position: {AI} Agents Need Authenticated Delegation},
  author={Tobin South and Samuele Marro and Thomas Hardjono and Robert Mahari and Cedric Deslandes Whitney and Alan Chan and Alex Pentland},
  booktitle={ICML},
  year={2025}
}

@article{yang2024formal,
  title={Formal mathematical reasoning: A new frontier in ai},
  author={Yang, Kaiyu and Poesia, Gabriel and He, Jingxuan and Li, Wenda and Lauter, Kristin and Chaudhuri, Swarat and Song, Dawn},
  journal={arXiv preprint arXiv:2412.16075},
  year={2024}
}

@inproceedings{andriushchenkojailbreaking,
  title={Jailbreaking Leading Safety-Aligned LLMs with Simple Adaptive Attacks},
  author={Andriushchenko, Maksym and Croce, Francesco and Flammarion, Nicolas},
  booktitle={ICLR},
  year={2025}
}

@misc{zod,
  title={Zod: Intro},
  author={Zod},
  year={2025},
  note={\url{https://zod.dev/}}
}

@inproceedings{betley2025emergent,
title={Emergent Misalignment: Narrow finetuning can produce broadly misaligned {LLM}s},
author={Jan Betley and Daniel Chee Hian Tan and Niels Warncke and Anna Sztyber-Betley and Xuchan Bao and Mart{\'\i}n Soto and Nathan Labenz and Owain Evans},
booktitle={Forty-second International Conference on Machine Learning},
year={2025}
}

@article{zhou2024robust,
  title={Robust prompt optimization for defending language models against jailbreaking attacks},
  author={Zhou, Andy and Li, Bo and Wang, Haohan},
  journal={Advances in Neural Information Processing Systems},
  volume={37},
  pages={40184--40211},
  year={2024}
}

@article{robey2025smoothllm,
title={Smooth{LLM}: Defending Large Language Models Against Jailbreaking Attacks},
author={Alexander Robey and Eric Wong and Hamed Hassani and George J. Pappas},
journal={Transactions on Machine Learning Research},
issn={2835-8856},
year={2025}
}

@article{nasr2025attacker,
  title={The attacker moves second: Stronger adaptive attacks bypass defenses against LLM jailbreaks and prompt injections},
  author={Nasr, Milad and Carlini, Nicholas and Sitawarin, Chawin and Schulhoff, Sander V and Hayes, Jamie and Ilie, Michael and Pluto, Juliette and Song, Shuang and Chaudhari, Harsh and Shumailov, Ilia and others},
  journal={arXiv preprint arXiv:2510.09023},
  year={2025}
}

@article{shvartzshnaider2025position,
  title={Position: Contextual integrity is inadequately applied to language models},
  author={Shvartzshnaider, Yan and Duddu, Vasisht},
  journal={arXiv preprint arXiv:2501.19173},
  year={2025}
}

@incollection{sandhu1998role,
  title={Role-based access control},
  author={Sandhu, Ravi S},
  booktitle={Advances in computers},
  volume={46},
  pages={237--286},
  year={1998},
  publisher={Elsevier}
}

\end{document}